\documentclass{aa}  
\usepackage{graphicx}
\usepackage{amsmath}
\usepackage{amssymb}
\usepackage{txfonts}
\usepackage{xcolor}

\newcommand{\rosl}{Roentgen Satellit}
\newcommand{\ros}{ROSAT}

\newcommand{\xmm}{{\it XMM-Newton}}

\newcommand{\eros}{eROSITA}
\newcommand{\erosl}{{extended Roentgen Survey with an Imaging Telescope Array}}
\newcommand{\srg}{Spectrum Roentgen Gamma}

\newcommand{\sxrt}{{\it Swift}/XRT}

\newcommand{\nh}{N_{\rm H}}
\newcommand{\gaia}{{\it Gaia}}

\def \magzs{\object{RX~J0720.4--3125}}

\def \magot{\object{RX~J1308.6+2127}}

\def \carINS{\object{2XMM~J104608.7--594306}}

\def \zsfs{\object{J06571}}
\def \otos{\object{J13171}}
\def \tmzsfs{\object{PSR~B0656+14}}

\def \ztto{\object{4XMM J022141.5--735632}}
\def \fluxcgs{erg~s$^{-1}$~cm$^{-2}$}
\def \jotos{\object{eRASSU~J131716.9--402647}}
\def \jzsfs{\object{eRASSU~J065715.3+260428}}

\def \rigi{\object{4XMM~J031722.7--663704}}
\def \rigii{\object{4XMM~J175437.8--294149}}
\def \rigiii{\object{4XMM~J180528.2--273158}}

\def \demi{\object{2RXS~J051541.7+010528}}
\def \demii{\object{2RXS~J092055.5+175526}}
\def \demiii{\object{2RXS~J101449.9+004101}}

\def \catvoton{\object{V1309~Ori}}

\begin{document} 

%

\title{Thermally emitting isolated neutron star candidates from the SRG/eROSITA All-Sky Survey}

\author{J.~Kurpas\inst{1,2}
   \and A.~D.~Schwope\inst{1}
   \and A.~M.~Pires\inst{1,3}
   \and F.~Haberl\inst{4}
}
\offprints{J. Kurpas}
\institute{Leibniz-Institut f\"ur Astrophysik Potsdam (AIP), An der Sternwarte 16, 14482 Potsdam, Germany
   \email{jkurpas@aip.de} 
   \and
   Potsdam University, Institute for Physics and Astronomy, Karl-Liebknecht-Stra\ss e 24/25, 14476 Potsdam, Germany
   \and
   Center for Lunar and Planetary Sciences, Institute of Geochemistry, Chinese Academy of Sciences, 99 West Lincheng Rd., 550051 Guiyang, China
   \and
   Max-Planck-Institut f\"ur extraterrestrische Physik, Giessenbachstra\ss e 1, 85748 Garching, Germany
}
\date{Received ...; accepted ...}
\titlerunning{Thermal isolated neutron star candidates from eROSITA}
\authorrunning{J.~Kurpas et al.}

\abstract
{The SRG/eROSITA All-Sky Survey (eRASS) allows for the creation of a complete sample of X-ray dim isolated neutron stars (XDINSs), which will significantly facilitate the study of their population properties, evolution, and connection to other families of isolated neutron stars (INSs). In this work, we conduct a systematic search for XDINSs on the eROSITA western Galactic hemisphere down to an X-ray flux limit of 10$^{-13}$~\fluxcgs\ and discuss the resulting candidate sample. Consistently with the properties of the known XDINSs, we selected all eRASS sources possessing a soft X-ray spectral distribution and that are unlikely to be associated with optical or infrared sources. Our selection criteria allowed us to recover all known XDINSs and previously proposed candidates. In addition, we put forward 33 new candidate members for dedicated follow-up identification campaigns. We found the resulting candidate sample to be about 30 -- 50\% complete, mainly due to source confusion and the stringent cross-matching criteria adopted to select the most promising candidates for immediate follow-up investigation. The candidates of the sample presented here can be broadly  divided into two groups: 13 rather soft and 20 hot and somewhat hard X-ray emitters. Interestingly, the remarkably thermal nature of the candidates in the first group as well as their spatial distribution, lack of known counterparts, and absence of significant flux variability agree well with the properties of other confirmed thermally emitting INSs. For the candidates in the second group, the current observational data do not allow one to discern between rotation-powered or recycled pulsars, cataclysmic variables, or quiescent neutron stars in binary systems or even to rule out an extragalactic nature. On the basis of population synthesis and the estimated source completeness of the search, we expect that between one and three new XDINSs are among the already singled-out list of XDINS candidates -- a long-sought increase in the proposed number of members of this elusive class of X-ray emitters.}

\maketitle
\section{Introduction\label{sec_intro}}

Thermal emission originating from the surface of isolated neutron stars (INSs) has been observed for a broad range of different INS types. However, the total number of INSs for which thermal emission has been detected is still low with respect to the overall known population \citep[e.g.][]{2020MNRAS.496.5052P}. Studying the thermal emission of INSs allows for investigation into the physical state of the matter, composition, and emission processes of the neutron star atmosphere \citep[e.g.][]{2002nsps.conf..263Z}; calibration of evolutionary models and cooling curves \citep[e.g.][]{2021ApJ...914..118D}; and estimation of the neutron star radius, a requirement to constrain the equation of state of cold nuclear matter \cite[e.g.][]{2016ARA&A..54..401O}. Thus, the identification of new thermally emitting INSs is of very high interest.

The cleanest thermal INS spectra that can be observed are those from the class of X-ray dim isolated neutron stars \cite[XDINSs;][]{2009ASSL..357..141T}. In comparison to most other INS types \citep[see e.g.][and references therein]{2023Univ....9..273P}, these objects possess long spin periods (3 -- 17~s), comparably strong magnetic fields ($10^{13}-10^{14}$~G), and thermal luminosities that exceed the observed loss of rotational energy, which suggest an evolution governed by the magnetic field and a possible connection with magnetars \citep{2013MNRAS.434..123V}. Evidence for the presence of non-thermal emission components has only recently been discovered in the stacked X-ray spectra of some of the XDINSs \citep{2017PASJ...69...50Y, 2019PASJ...71...17Y, 2020ApJ...904...42D, 2022MNRAS.516.4932D}. The absence of supernova remnants, pulsar wind nebulae, and coherent pulsed radio and gamma-ray emission as well as faint counterparts at optical or ultraviolet wavelengths make XDINSs particularly difficult to detect. The seven known XDINSs were originally identified from observations at soft X-ray energies with the \rosl\ \citep[\ros;][]{2007Ap&SS.308..181H}, where thermal emission is most prevalent. The seven XDINSs are all X-ray bright \citep[implying large X-ray-to-optical flux ratios of $\log\left(f_X/f_\mathrm{opt}\right)\gtrsim4$;][]{2004AdSpR..33..638H,2008AIPC..983..331K}, lowly absorbed, and situated in the immediate solar vicinity \citep[$<1~$kpc; see][and references therein]{2009A&A...497..423M}, which implies a local birthrate that is akin to that of the much more numerous class of rotation-powered pulsars \citep[RPPs;][]{2009ASSL..357...91B}. As such, XDINSs may constitute a substantial part of the Galactic INS population \citep{2008MNRAS.391.2009K}. Since their initial discovery in the 1990s, several efforts to identify new members have been carried out by making use of large serendipitous datasets from pointed X-ray observations \citep{2009A&A...504..185P, 2022MNRAS.509.1217R, 2022A&A...666A.148P} or searches in the data of the ROSAT All-Sky Survey \citep{2008ApJ...672.1137R, 2011AJ....141..176A, 2024ApJ...961...36D}. These campaigns have led to a growing list of previously unknown soft X-ray emitters and INS candidates, though an XDINS nature is yet to be established for them.

From December 2019 to February 2022, the \erosl\ (\eros) instrument aboard the \srg\ {(SRG)} mission conducted 4.3 entire scans of the sky in the X-ray band (between 200~eV and 8~keV) as part of the eROSITA All-Sky Survey \cite[eRASS;][]{2021A&A...647A...1P,2024A&A...682A..34M}. In comparison to its predecessor, the ROSAT All-Sky Survey \citep[][]{1999A&A...349..389V,2016A&A...588A.103B}, \eros\ offers a much improved sensitivity at soft X-ray energies \cite[][]{2024A&A...682A..34M} and better positional accuracy. Given the large sky coverage, eRASS is well suited to search for new thermally emitting INSs, in particular fainter XDINSs \citep{2017AN....338..213P,2021ARep...65..615K}. We searched the eRASS source catalogues \citep[see][]{2024A&A...682A..34M} to identify suitable INS candidates and initiated a dedicated follow-up campaign at X-ray and optical energies to confirm their nature. The overall properties of the first two proposed candidates, \jzsfs\ and \jotos, which have been singled out on the basis of extremely high X-ray-to-optical flux ratios, predominantly soft and constant X-ray emission, and absence of catalogued counterparts, were presented in \cite{2023A&A...674A.155K}. Additional observations of \jotos\, unveiling the spin period of the neutron star and the presence of possible spectral absorption lines, suggest that the properties of the X-ray source are remarkably similar to those of a highly magnetised INS, an XDINS, or a high magnetic field pulsar \cite[][]{2024A&A...683A.164K}.

In this work, we present a detailed account of our search procedure and the overall properties of the resulting sample of INS candidates. The results of still on-going follow-up identification campaigns will be presented elsewhere. In Sect.~\ref{sec_obs}, we start by outlining the selection criteria, then proceed to present the reduction steps involved in the analysis of the \eros\ data and how the optical and infrared (IR) magnitude limits were derived. In Sect.~3, we estimate the completeness and false-deselection rate of our search, compare the selection to previous searches, and present the results of the spectral analysis, the inferred X-ray variability, and the current X-ray-to-optical flux ratios of our sample of candidates. In Sect.~4, we discuss the properties of the identified candidates in the context of the known population of INSs, particularly the class of XDINSs. Our conclusions and outlook to further improve future searches, especially based on \eros\ data, are presented in Sect.~\ref{sec_concl}.

\section{Data reduction\label{sec_obs}}

\subsection{Candidate selection}

The eROSITA instrument has conducted four complete sky surveys and initiated the fifth. Therefore, for each sky location and detected X-ray source, X-ray information from four or five epochs, roughly obtained six months apart, are available. To identify INS candidates, we followed the procedure briefly outlined in \cite{2023A&A...674A.155K}. An overview of all applied cuts and the remaining number of candidates is given in Table~\ref{tab_selsteps}. The starting point of the candidate selection is a stacked \eros\ source catalogue\footnote{The source catalogue was compiled based on pipeline version c020 and shared internally among the members of the German \eros\ consortium.} containing sources that were detected using all the available observations. We required well detected and point-like sources (detection likelihood $>10$ at energies 0.2 -- 2.3~keV and extend = 0\arcsec) and that are brighter than a limiting flux of $10^{-13}$~\fluxcgs\ (within $1\sigma$; 0.2 -- 2~keV). This results in an input sample of $102,620$ sources. At the chosen flux limit, a follow-up program to confirm or reject the INS nature is feasible with a moderate amount of observing resources. Accordingly, population synthesis models predict about $26\pm4$ INSs to be discovered all-sky at this flux level \citep{2017AN....338..213P}.

\begin{figure}[t]
\begin{center}
\includegraphics[width=\linewidth]{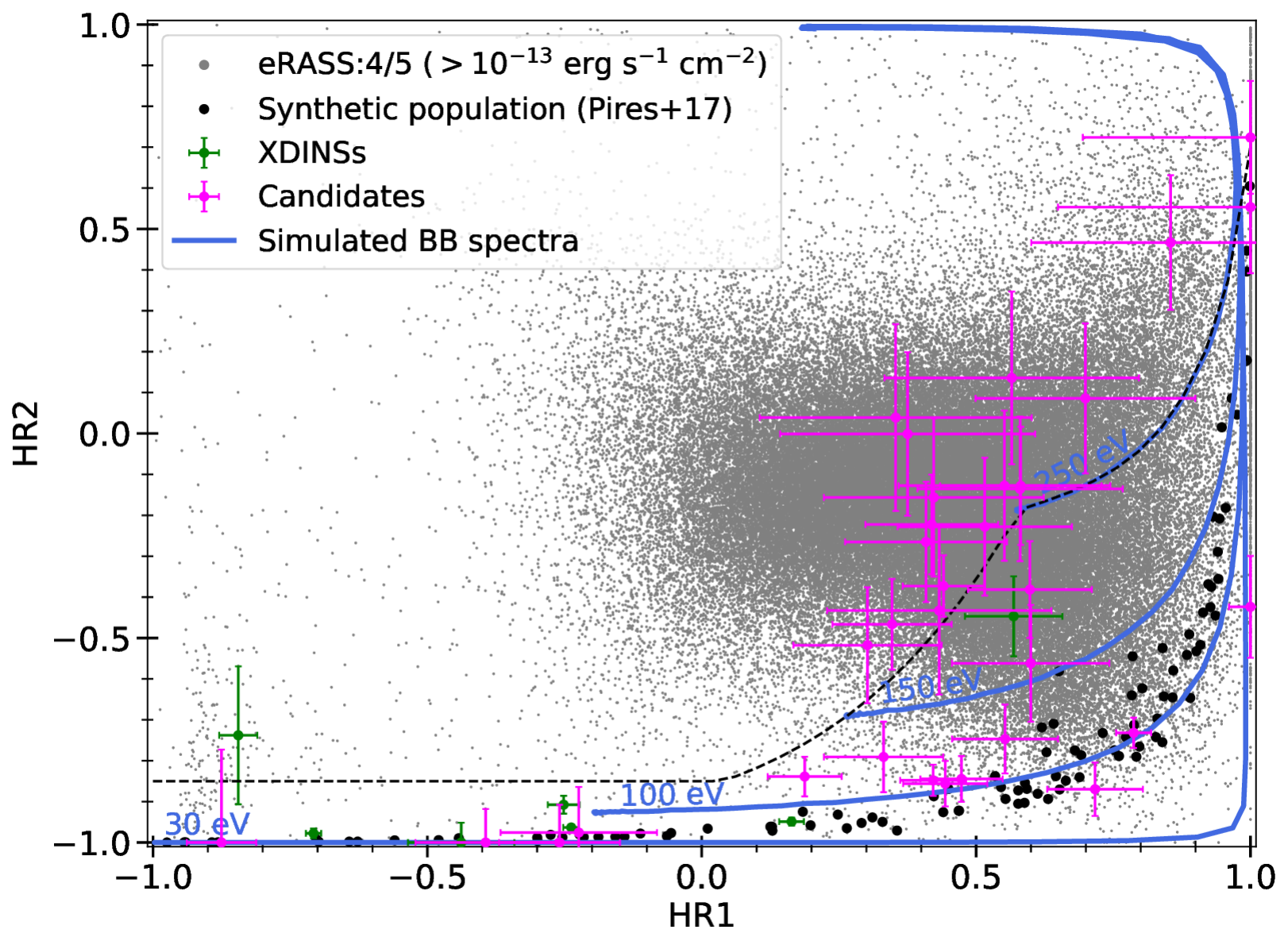}\vskip1pt
\includegraphics[width=\linewidth]{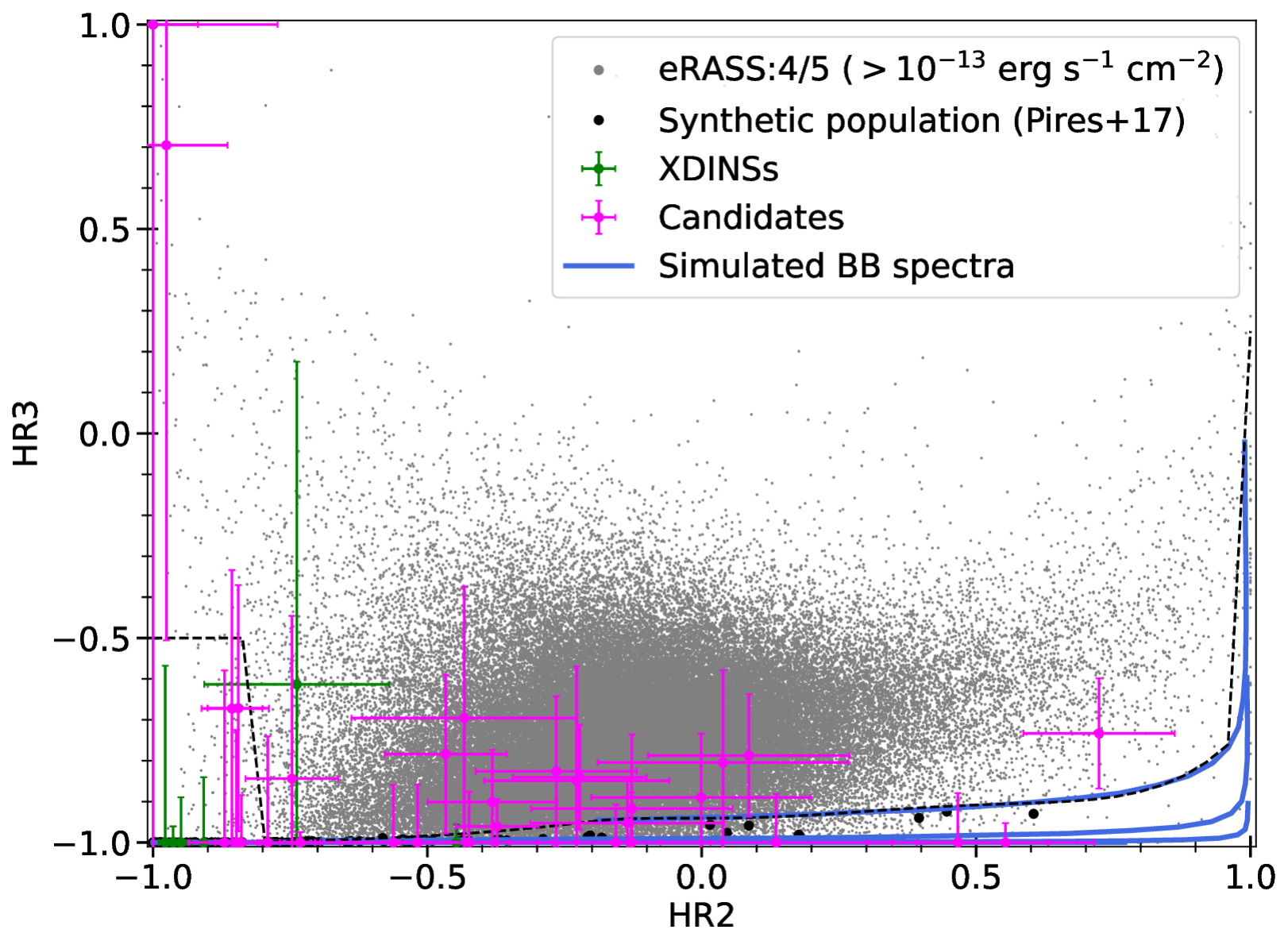}
\end{center}
\caption{Hardness ratio plots presenting the hardness ratio distribution for all the selected candidates (magenta), the five confirmed XDINSs on the western Galactic hemisphere and recently proposed candidates \citep[green;][]{2015A&A...583A.117P, 2022A&A...666A.148P}, and the synthetic INS population from \citet[black;][]{2017AN....338..213P}. The applied cut to select the soft X-ray emitting candidates from all the \eros\ sources above the applied flux limit (grey) is indicated by the black dashed line and is based on simulated black body spectra of varying temperature and absorption strengths (blue lines). (See the text for additional information on these simulations.)}
\label{fig_hr_plots}
\end{figure}

Based on cuts in hardness ratio ("X-ray colour") space, we only considered eRASS sources with an X-ray spectrum consistent with that of XDINSs, within 1$\sigma$. More specifically, the hardness ratio compares the number of counts in two neighboring X-ray energy bands, e.g. $\mathrm{HR} = \frac{\mathrm{CTS}_2-\mathrm{CTS}_1}{\mathrm{CTS}_1+\mathrm{CTS}_2}$, for bands 1 and 2. In the case of the eRASS catalogues, these were defined for energy bands 0.2 -- 0.5~keV, 0.5 -- 1~keV, 1 -- 2~keV, and 2 -- 5~keV. In order to identify the parameter space in hardness ratio where thermally emitting INSs are located, we expect their spectra to follow an absorbed blackbody continuum. As such, we used the latest eROSITA response files and the XSPEC \citep[version 12.12.0;][]{1996ASPC..101...17A} \texttt{fakeit} command to simulate absorbed blackbody models with temperature and hydrogen column density ranges of 30 -- 250~eV and 0 -- 10$^{23}$~cm$^{-2}$, respectively (see the diagrams in Fig.~\ref{fig_hr_plots}). The adopted range in temperature includes that of the observed XDINSs \citep[40 -- 100~eV;][]{2009ASSL..357..141T}, while leaving sufficient margin to select also younger and possibly hotter members of this class. Based on the simulation results, we defined cuts in the hardness ratio spaces $\mathrm{HR}_1 \times \mathrm{HR}_2$ and $\mathrm{HR}_2 \times \mathrm{HR}_3$ (black dashed lines in Fig.~\ref{fig_hr_plots}). Their application results in a sample of $\sim 27,670$ sources.

\begin{table}
\caption{Applied selection steps.\label{tab_selsteps}}
\centering

\begin{tabular}{llrrrrrrr}
\hline\hline
Step & Description & Remaining sources\\
\hline
0 & Initial catalogue & $\sim3\times 10^{6}$\\
1 & Well detected point sources & 102,620\\
2 & Hardness ratio cuts & 27,670\\
3 & Cross-match & 554\\
4 & Screening & 41\\
5 & Exclude known INSs & 33\\
\hline
\end{tabular}
\end{table}

The third selection cut in Table~\ref{tab_selsteps} is motivated by the large X-ray-to-optical flux ratio values that are observed for the known XDINSs. This property alone separates INSs from any other soft X-ray emitter and can be used to confirm the INS nature, but also to efficiently pinpoint promising candidates for follow-up investigations. Given the flux level of our list of INS candidates ($\sim (1 - 5)\times10^{-13}$~\fluxcgs), their expected optical and IR emission would be too faint \citep[e.g. $m_V>27$~mag under the assumption of $\log\left(f_X/f_\mathrm{opt}\right)\gtrsim4$ that is observed for the known population;][]{2004AdSpR..33..638H,2008AIPC..983..331K} to be detected in most wide-area photometric catalogues. Thus, we cross-correlated the \eros\ source positions against those of optical and IR objects present in \gaia\  DR3 \citep{2023A&A...674A...1G}, Legacy Survey DR10 \citep{2019AJ....157..168D}, Pan-STARRS DR1 \citep{2016arXiv161205560C}, DeCAPS2 \citep{2023ApJS..264...28S}, VISTA Hemisphere Survey \citep{2013Msngr.154...35M}, unWISE \citep{2019ApJS..240...30S}, and CatWISE2020 \citep{2021ApJS..253....8M}, adopting the multi-catalogue probabilistic method implemented in the ARCHES tool \citep{2017A&A...597A..89P}. The ARCHES tool computes individual probabilities for all possible combinations of X-ray, optical and IR sources, based on their angular distance, local distribution, and positional accuracy. To discard X-ray sources with likely optical and IR counterparts, we defined the probability that the X-ray source has no counterpart at all (hereafter the non-matching probability) as the lowest likelihood value of non-association among all the individual matching combinations computed by ARCHES. We then retained the 554 X-ray sources with non-matching probabilities above a threshold value of 50\%.

We finally screened the remaining sources to exclude spurious candidates. First, we required overall long-term flux stability between \eros\ visits. In practice, we excluded transient sources that were not individually detected in all the single eRASS scans\footnote{The candidate J08071, which otherwise survives the selection, was detected in an internal earlier version of the eRASS1 source catalogue, but is not included in the published version \citep{2024A&A...682A..34M}. The source is nonetheless reliably detected in all the other eRASS scans (detection likelihood values within 17 and 40). The non-detection in eRASS1 is not problematic since we found the eRASS1 flux upper limit of $1.38\times 10^{-13}$~\fluxcgs\ \citep[0.2 -- 2.3~keV;][]{2024A&A...682A..35T} to exceed the observed source flux.} since the eRASS is expected to be generally complete at the applied flux limit \citep[the completeness at a given flux and exposure time is exemplarily shown for the first \eros\ sky scan in Fig. 10 of][]{2024A&A...682A..34M}. Similarly, we found that some remaining sources showed a large spread between the single eRASS scan positions, to a degree that it was difficult to still associate these single eRASS scan positions with the detection in the stacked catalogue. Many of these objects were located in regions with increased X-ray sky background, like supernova remnants, making them more likely to be spurious detections and not real astronomical point sources. For a clean selection we discarded these sources as well. Finally, we found that the remaining sample still contained eRASS sources that can be associated with optically very bright counterparts (e.g. nearby stars with magnitudes $\lesssim14$~mag) that were also consequently removed. While the ARCHES cross-match should have originally associated these objects with their optical counterparts, all these sources possess very accurate optical source positions and a significant displacement of the eRASS sky position. This led to spuriously large non-matching probabilities. We included the proper motion into the ARCHES match such that an intrinsic shift in the sky position should not be the cause of the observed displacement. We conclude that for \eros\ sources in the vicinity of an optically bright object larger systematic uncertainties may need to be considered. The screening left us with a candidate sample of 41 sources. The published version of the eRASS1 source catalogue contains flags \citep[see][]{2024A&A...682A..34M}, marking objects that could be spurious or potentially affected by optical loading. We verified that none of the remaining targets are flagged in any way.

We used the SIMBAD database \citep{2000A&AS..143....9W} and the ATNF pulsar catalogue \citep[Version 2.0;][]{2005AJ....129.1993M} to identify already known neutron stars in the selected sample. In total, we found the sample to include eight archival neutron stars. These were the five known XDINSs in the western Galactic hemisphere, the two proposed XDINS candidates \carINS\ \citep{2015A&A...583A.117P} and \ztto\ \citep{2022MNRAS.509.1217R, 2022A&A...666A.148P} and the RPP \tmzsfs, which also emits strong thermal radiation at soft X-ray energies \citep[e.g.][]{2022A&A...661A..41S}. This left us with 33 previously unknown sources that have survived the selection. We list the X-ray sky position, Galactic latitude, angular distance to the next optical or IR source, probability of non-association, hardness ratios, and \eros\ exposure times of this sample in Table~\ref{tab_cands_prop}.

\begin{table*}
\caption{Properties of the selected candidates.
\label{tab_cands_prop}}
\centering
\scalebox{.65}{
\begin{tabular}{lrrrrrrrrrrrrrr}
\hline\hline
Source & RA & DEC & Error\tablefootmark{(a)} & b & P$_\text{no-match}$\tablefootmark{(b)} & Separation\tablefootmark{(c)} & HR1\tablefootmark{(d)} & HR2\tablefootmark{(d)} & HR3\tablefootmark{(d)} & CTS\tablefootmark{(e)} & T$_{exp}$\tablefootmark{(f)} & Magnitude\tablefootmark{(g)} & Filter\tablefootmark{(h)} & $f_X/f_{opt}$\tablefootmark{(i)} \\
eRASSU & [\degr] & [\degr] & [\arcsec] & [\degr] & [\%] &  [\arcsec] & & & & & [s] & limit [mag] & & \\
 \hline
J052152.0-503812 & 80.467 & -50.6369 & 1.16 & -34.58 & 100 & 5.81\arcsec(5.01$\sigma$) &  0.44(0.08) &  -0.37(0.08) &  -0.96(0.07) &  232 & 3226 & 25.48 & r &            449  \\
J062748.5-260240 & 96.9523 & -26.0446 & 1.50 & -16.41 & 60 & 2.99\arcsec(2.00$\sigma$) &  0.3(0.14) &  -0.52(0.15) &  -1.0(0.15) &  64 & 964 & 21.73 & r &       14   \\
J065715.3+260428\tablefootmark{(j)} & 104.314 & 26.0746 & 1.31 & 12.75 & 100 & 15.15\arcsec(11.6$\sigma$) &  0.33(0.11) &  -0.79(0.09) &  -1.0(0.26) &  88 & 659 & 22.21 & r &        859  \\
J071828.7+141044 & 109.62 & 14.1789 & 2.05 & 12.31 & 84 & 5.32\arcsec(2.59$\sigma$) &  1.0(0.4) &  0.55(0.17) &  -1.0(0.05) &  27 & 630 & 22.16 & r &       118  \\
J072302.3-194225 & 110.7599 & -19.7071 & 1.62 & -2.21 & 83 & 5.50\arcsec(3.40$\sigma$) &  -0.22(0.15) &  -0.98(0.12) &  0.7(1.3) &  53 & 764 & 21.93 & r &        22   \\
J072635.8-001642 & 111.6496 & -0.2785 & 2.52 & 7.69 & 91 & 6.24\arcsec(2.48$\sigma$) &  0.85(0.26) &  0.47(0.17) &  -1.0(0.13) &  36 & 652 & 21.92 & r &        24   \\
J080701.3+023823 & 121.7558 & 2.6398 & 1.76 & 17.99 & 54 & 6.17\arcsec(3.51$\sigma$) &  0.52(0.16) &  -0.23(0.17) &  -0.85(0.28) &  47 & 598 & 24.50 & r &       512  \\
J080717.2-182019 & 121.8217 & -18.3387 & 1.95 & 7.53 & 52 & 2.90\arcsec(1.49$\sigma$) &  0.6(0.15) &  -0.56(0.15) &  -1.0(0.15) &  47 & 697 & 22.00 & r &        17   \\
J081952.1-131930 & 124.9672 & -13.3251 & 1.38 & 12.76 & 100 & 8.56\arcsec(6.21$\sigma$) &  -0.88(0.07) &  -1.0(0.23) &  1.0(5.0) &  78 & 677 & 22.99 & r &       146  \\
J083039.8+353102 & 127.6661 & 35.5172 & 1.96 & 34.67 & 100 & 7.79\arcsec(3.98$\sigma$) &  0.56(0.24) &  0.14(0.22) &  -1.0(0.12) &  29 & 574 & 24.12 & r &       263  \\
J084046.2-115222 & 130.1926 & -11.873 & 1.65 & 17.82 & 79 & 7.18\arcsec(4.34$\sigma$) &  -0.39(0.13) &  -1.0(0.09) &  1.0(2.7) &  51 & 622 & 22.12 & r &        36   \\
J085257.5+270013 & 133.2399 & 27.0039 & 1.92 & 37.45 & 100 & 9.45\arcsec(4.91$\sigma$) &  0.58(0.19) &  -0.14(0.18) &  -0.95(0.11) &  40 & 577 & 24.56 & r &      425  \\
J085351.2+272536 & 133.4634 & 27.4268 & 2.42 & 37.75 & 86 & 7.57\arcsec(3.13$\sigma$) &  1.0(0.4) &  0.72(0.14) &  -0.73(0.14) &  32 & 585 & 24.51 & r &      1217 \\
J090421.1-383956 & 136.0882 & -38.6656 & 1.60 & 5.50 & 100 & 7.09\arcsec(4.44$\sigma$) &  1.0(0.04) &  -0.42(0.13) &  -1.0(0.13) &  63 & 919 & 23.64 & r &        67   \\
J093317.6-251252 & 143.3236 & -25.2147 & 1.28 & 19.17 & 62 & 3.99\arcsec(3.13$\sigma$) &  0.42(0.13) &  -0.22(0.13) &  -0.84(0.13) &  94 & 695 & 22.35 & r &        75   \\
J100957.4+125914 & 152.4893 & 12.9873 & 2.92 & 49.75 & 50 & 6.71\arcsec(2.30$\sigma$) &  0.43(0.21) &  -0.43(0.21) &  -0.7(0.4) &  29 & 525 & 24.62 & r &       154  \\
J101110.9+083329 & 152.7957 & 8.5582 & 1.86 & 47.85 & 58 & 6.43\arcsec(3.45$\sigma$) &  0.42(0.21) &  -0.16(0.2) &  -1.0(0.1) &  36 & 542 & 24.63 & r &       223  \\
J103951.1+323442 & 159.9631 & 32.5785 & 2.35 & 61.07 & 100 & 11.40\arcsec(4.86$\sigma$) &  0.35(0.25) &  0.04(0.23) &  -0.8(0.23) &  27 & 609 & 24.35 & r &       338  \\
J110240.2-250706 & 165.6678 & -25.1184 & 1.33 & 31.55 & 99 & 5.17\arcsec(3.88$\sigma$) &  0.35(0.11) &  -0.47(0.12) &  -0.78(0.2) &  104 & 815 & 24.65 & r &       606  \\
J113351.2-522712 & 173.4634 & -52.4535 & 1.18 & 8.63 & 80 & 2.96\arcsec(2.51$\sigma$) &  0.55(0.1) &  -0.75(0.09) &  -0.8(0.4) &  99 & 1672 & 23.37 & r &        55   \\
J121826.7+071144 & 184.6113 & 7.1956 & 2.28 & 68.59 & 56 & 4.52\arcsec(1.98$\sigma$) &  0.38(0.24) &  0.0(0.2) &  -0.89(0.16) &  35 & 715 & 24.69 & r &       170  \\
J122150.6-193438 & 185.4611 & -19.5774 & 1.40 & 42.75 & 99 & 5.31\arcsec(3.79$\sigma$) &  0.6(0.12) &  -0.38(0.12) &  -0.9(0.13) &  85 & 1083 & 24.48 & r &       207  \\
J125214.5-591759 & 193.0606 & -59.2999 & 1.04 & 3.57 & 85 & 0.59\arcsec(0.57$\sigma$) &  0.47(0.08) &  -0.84(0.06) &  -0.7(0.4) &  164 & 1802 & 23.32 & r &        77   \\
J130406.1-533816 & 196.0256 & -53.6379 & 1.13 & 9.19 & 99 & 3.75\arcsec(3.32$\sigma$) &  0.44(0.08) &  -0.86(0.06) &  -0.7(0.4) &  147 & 1735 & 23.24 & r &        63   \\
J130821.3+130319 & 197.0891 & 13.0555 & 1.69 & 75.38 & 100 & 7.42\arcsec(4.39$\sigma$) &  0.58(0.15) &  -0.13(0.15) &  -1.0(0.13) &  75 & 1060 & 24.63 & r &       206  \\
J131716.9-402647\tablefootmark{(j)} & 199.3205 & -40.4464 & 0.83 & 22.16 & 100 & 5.16\arcsec(6.21$\sigma$) &  0.42(0.07) &  -0.85(0.04) &  -1.0(0.28) &  415 & 1464 & 24.88 & r &       10801\\
J133355.8-665647 & 203.4828 & -66.9466 & 0.81 & -4.42 & 65 & 1.64\arcsec(2.03$\sigma$) &  0.79(0.04) &  -0.73(0.04) &  -1.0(0.026) &  468 & 1836 & 23.31 & r &       188  \\
J134725.4-363415 & 206.8561 & -36.5709 & 1.31 & 24.95 & 99 & 4.50\arcsec(3.42$\sigma$) &  -0.26(0.12) &  -1.0(0.1) &   &  100 & 1376 & 25.00 & r &       381  \\
J174858.8-510951 & 267.245 & -51.1642 & 0.92 & -11.83 & 100 & 5.60\arcsec(6.10$\sigma$) &  0.19(0.07) &  -0.84(0.05) &  -1.0(0.12) &  256 & 920 & 21.21 & g &        31   \\
J175415.0-740127 & 268.5625 & -74.0244 & 1.58 & -22.20 & 84 & 4.43\arcsec(2.81$\sigma$) &  0.41(0.15) &  -0.27(0.15) &  -0.83(0.19) &  66 & 1198 & 24.84 & r &       194  \\
J175532.5-503748 & 268.8858 & -50.6301 & 1.47 & -12.48 & 77 & 4.89\arcsec(3.33$\sigma$) &  0.72(0.09) &  -0.87(0.07) &  -1.0(0.5) &  82 & 846 & 21.33 & g &        11   \\
J183349.6-353707 & 278.4569 & -35.6188 & 2.18 & -12.12 & 56 & 4.49\arcsec(2.06$\sigma$) &  0.55(0.2) &  -0.13(0.19) &  -0.92(0.19) &  42 & 666 & 20.89 & g &         17   \\
J214920.9-443050 & 327.3374 & -44.514 & 1.77 & -49.85 & 53 & 4.29\arcsec(2.42$\sigma$) &  0.7(0.21) &  0.09(0.19) &  -0.79(0.15) &  38 & 673 & 25.24 & r &      298  \\
\hline
\end{tabular} }
\tablefoot{During the observations, eROSITA was operated in SURVEY mode.
\tablefoottext{a}{$1\sigma$ positional accuracy. Calculated as described in Table~2 of \cite{2023A&A...674A.155K}.}
\tablefoottext{b}{Probability from the ARCHES cross-match that the eROSITA source has no counterpart in any of the wide area optical or IR catalogues.}
\tablefoottext{c}{Angular distance to the closest optical or IR source.}
\tablefoottext{d}{Hardness ratio values. HR1 is computed between the 0.2 -- 0.5~keV and 0.5 -- 1~keV bands, HR2 between the 0.5 -- 1~keV and 1 -- 2~keV bands, and HR3 between the 1 -- 2~keV and 2 -- 5~keV bands.}
\tablefoottext{e}{Number of detected counts in the 0.2 -- 2~keV band.}
\tablefoottext{f}{Total CCD live time, uncorrected for vignetting effects.}
\tablefoottext{g}{Local magnitude limit computed from neighboring optical sources.}
\tablefoottext{h}{Passband of the local magnitude limit. The term $r$ refers to the Pan-STARRS, Legacy Survey, or DeCAPS2 $r$-band, whereas $g$ marks magnitudes in the \gaia\  $g$-band. All magnitudes are given in the AB system.}
\tablefoottext{i}{The X-ray-to-optical flux ratio, in relation to the $V$-band, was computed via $\log(f_\mathrm{X}/f_\mathrm{opt}) = \log(\frac{F_\mathrm{X}}{\mathrm{erg}\,\mathrm{s}^{-1}\mathrm{cm}^{-2}}) +\frac{m}{2.5}+5.39$, with $F_\mathrm{X}$ as the X-ray flux and $m$ the local magnitude limit. See Sect.~\ref{sec_fxfopt} for more details.}
\tablefoottext{j}{We list the derived properties for J06571 and J13171 also in this table, although they were already published in \cite{2023A&A...674A.155K} and \cite{2024A&A...683A.164K}.}
}
\end{table*}
%

\subsection{\eros\ data reduction}

The eRASS is conducted with the \eros\ instrument aboard the SRG mission \citep{2021A&A...647A...1P}. This instrument contains seven individual Wolter-I X-ray telescopes (dubbed TM1, TM2, ..., TM7; together forming a virtual telescope TM0) that simultaneously observe the sky. While the selection of XDINS candidates was done using the standardised pipeline-processed source catalogues \citep[][]{2022A&A...661A...1B}, directly studying the individual eROSITA spectra of the 33 selected candidates allowed for better characterisation of the sources. We retrieved the event lists corresponding to each of the 33 XDINS candidates' sky tiles in all available eRASS scans and analysed the events with the eROSITA Science Analysis Software System \citep[eSASS; user version 211214.0.5;][]{2022A&A...661A...1B}. The event lists, with photons from all telescope modules (TM0), were processed with the latest pipeline version c020.

\begin{figure}[t]
\begin{center}
\includegraphics[width=\linewidth]{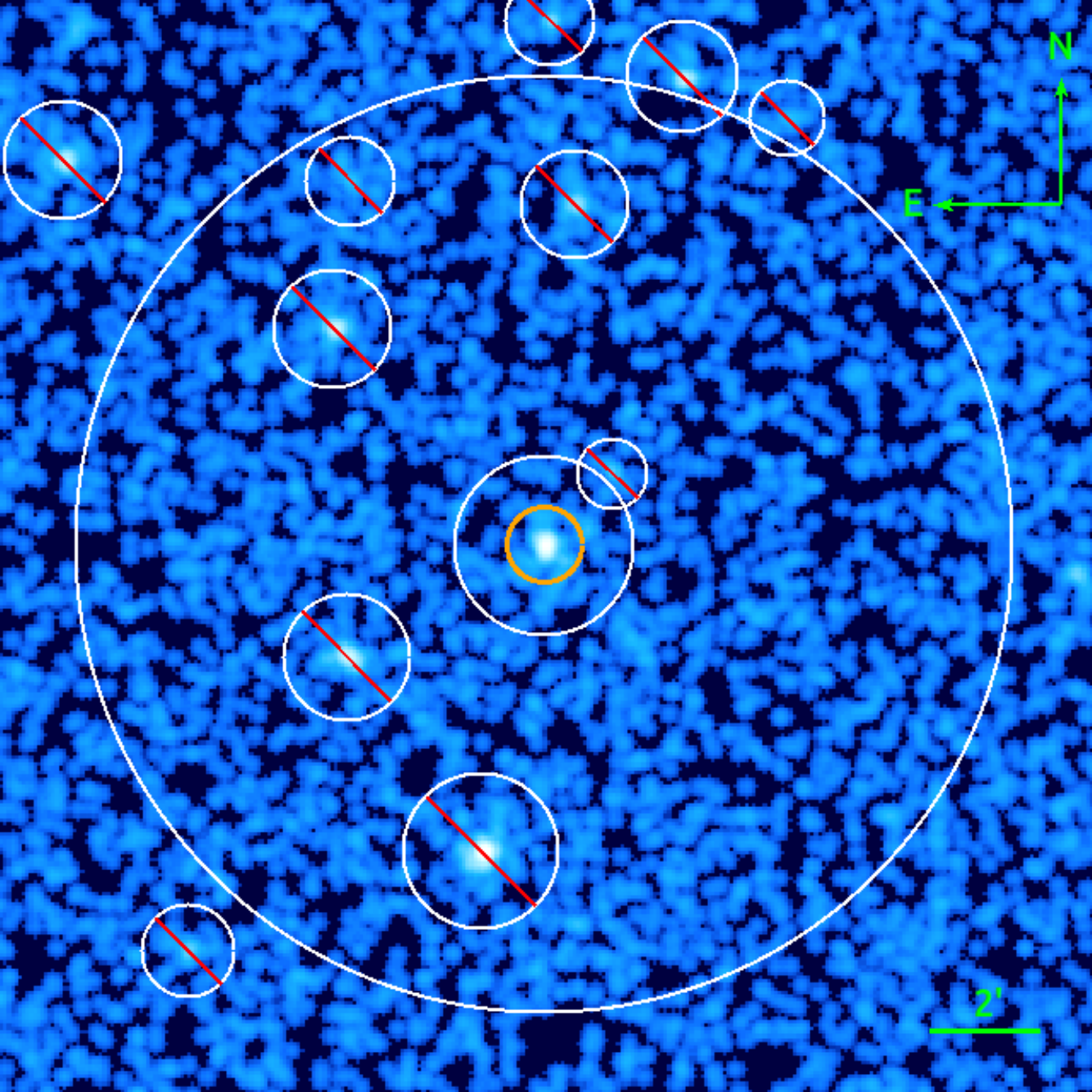}
\end{center}
\caption{\eros\ image of the field of J13472. The source extraction region is indicated by the orange circle, whereas the background region annulus and excluded field sources are marked by white circles.}
\label{fig_ero_img}
\end{figure}

For the spectral analysis, we adopted a region of interest of 15\arcmin\ around the position of each candidate and considered all valid photon patterns with good quality flags (PATTERN $\leq 15$ and FLAG=0xE000F000\footnote{The applied flag removes events that are corrupted, originate from a corrupted frame, originate from an artefact, were registered outside the field-of-view, or are located on a bad or dead pixel.}). Periods of high background were excluded by constructing updated good-time-interval (GTI) lists with the \texttt{flaregti} task. The remaining total net exposures and counts are presented in Table~\ref{tab_cands_prop} for all the sources. We then followed \citet[][]{2022A&A...661A...1B} to perform source detection in the 0.2 -- 2.3~keV band to locate the target and nearby field sources. The \texttt{srctool} task was then applied to produce optimised source and background regions based on the results of the source detection (see Fig.~\ref{fig_ero_img} for an example) and to extract the source and background spectra. The spectra were then grouped requiring at least one count per spectral bin with the FTOOLS task \texttt{grppha} \citep[version: 3.1.0;][]{1995ASPC...77..367B}.

\subsection{Estimation of the optical magnitude limits}

\begin{figure}[t]
\begin{center}
\includegraphics[width=\linewidth]{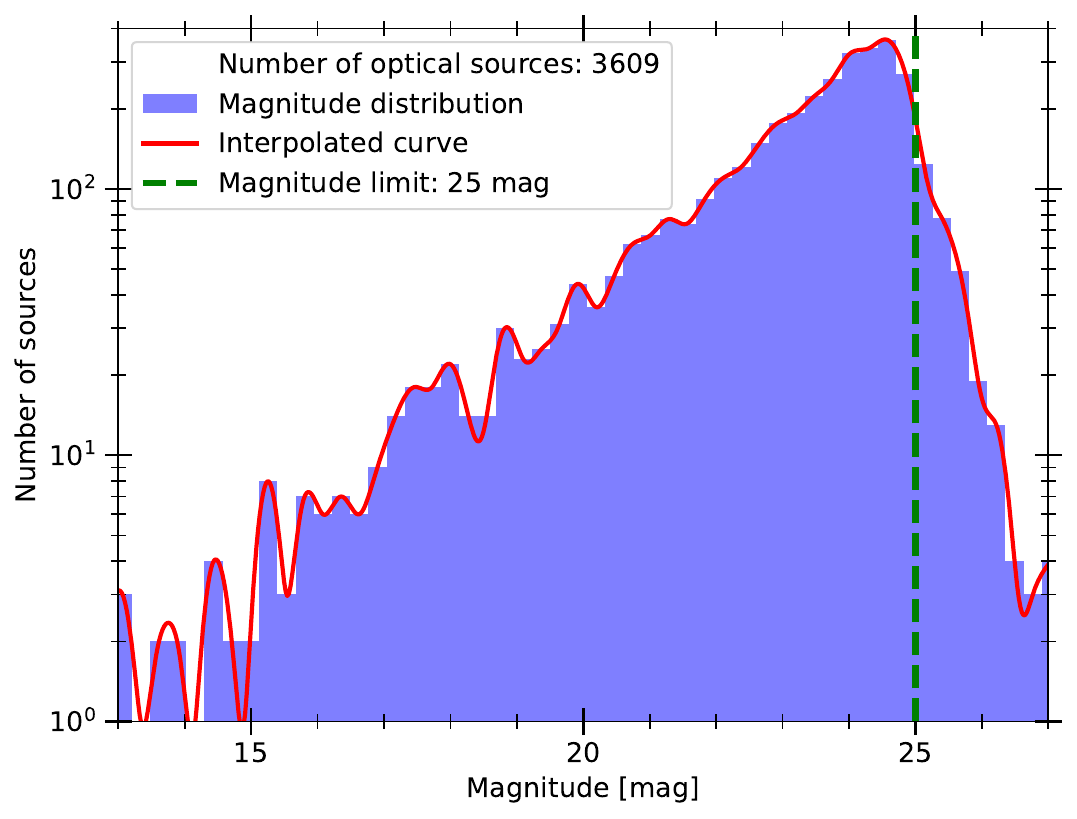}
\end{center}
\caption{Magnitude distribution of Legacy Survey DR10 sources in a 5\arcmin\ radius around the candidate J13472. The distribution was interpolated with a B-spline (red line), and the resulting local magnitude limit at 25~mag is indicated by the vertical green dashed line.}
\label{fig_mag_lim_est}
\end{figure}

We computed the local optical detection limit in the field of each XDINS candidate by collecting the magnitude values of all optical objects catalogued in the \gaia\  DR3, Pan-STARRS DR1, Legacy Survey DR10 and DeCAPS2 surveys within a search radius of 5\arcmin. The limiting magnitude was then defined as the value where the count number of optical objects in the distribution of magnitude values has decreased to half of the maximum of the histogram. For improved accuracy, for each field and available catalogue we used the \texttt{splrep} function, built into the scipy python package \citep{2020SciPy-NMeth}, to interpolate between the discrete histogram bins of the distribution. This approach is exemplarily presented in Fig.~\ref{fig_mag_lim_est} for the candidate J13472 (in the following, we will refer to individual candidates based on the first five digits of their position). The computed magnitude limits are noted for each candidate in Table~\ref{tab_cands_prop}.

\section{Results\label{sec_analysis}}
\subsection{Completeness and false-deselection fraction of the candidate selection}
In the following, we assess the efficiency of the method at selecting XDINS candidates from the eRASS catalogues based on the completeness and reliability of the sample. The long-term goal of this project is to enable the creation of a complete flux-limited sample of well-characterised XDINSs that will improve studies of their population properties and evolution. For this reason, the applied cut in X-ray flux at a limit of $10^{-13}$~\fluxcgs\ is justified since the inclusion of fainter sources in the envisaged sample is made difficult by the amount of needed follow-up to fully characterise their nature. Considering 50 simulations of the population synthesis model described in \citet{2017AN....338..213P}, the applied flux cut would keep ($46 \pm 13$)\% of the population expected to be in the eRASS catalogue version that we analysed\footnote{Adopting conservative optical blocking filter options \citep[see][for details]{2017AN....338..213P}}.

In principle, the applied hardness-ratio cuts could deselect good candidates. To prevent this, the cut was designed to be conservative by allowing for a much wider temperature (up to 250~eV) and hydrogen column density (up to $10^{23}$~cm$^{-2}$) range than is actually observed in the known XDINS population ($kT\sim40-100$~eV and $\nh <10^{21}$~cm$^{-2}$). Indeed, a comparison of these cuts with the simulated population of INSs in \citet{2017AN....338..213P} shows that we select all of them (see black dots in Fig.~\ref{fig_hr_plots}). Consequently, we expect that no good candidates were lost.

The selection step that is most prone to removing good candidates is the cross-matching of the \eros\ and archival optical or IR imaging catalogues. For an X-ray survey mission, \eros\ has unprecedentedly good positional accuracy (the 95\% $1\sigma$ positional error percentile is located at 2.6\arcsec, estimated from all soft sources surviving the hardness ratio cuts). However, a neutron star can erroneously be removed just by being located angularly close to an optical or IR source on the sky. Source confusion is expected to worsen in crowded sky regions, such as at low Galactic latitudes, due to the increased chance of assigning a spurious identification to a given X-ray source.

\begin{figure}[t]
\begin{center}
\includegraphics[width=\linewidth]{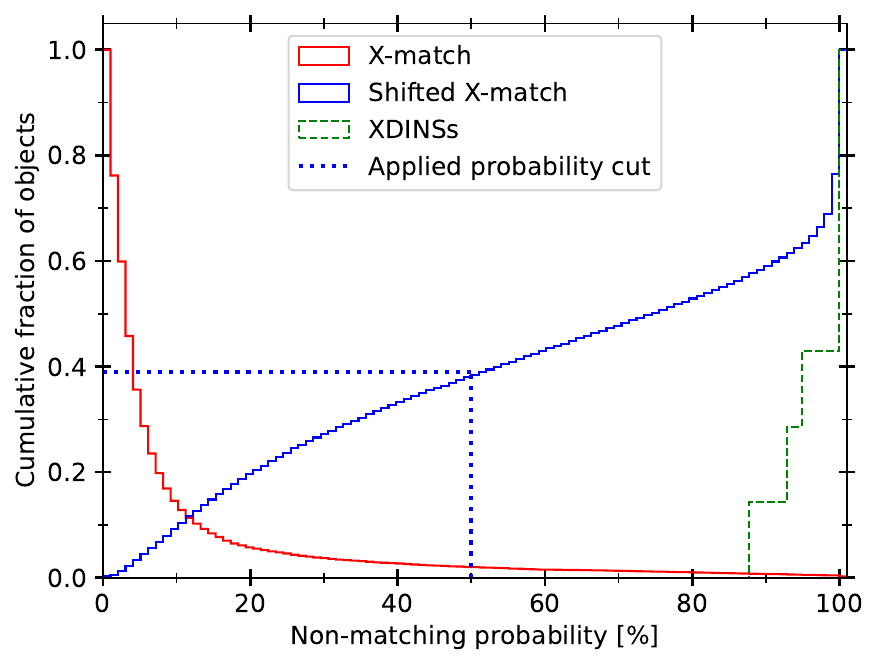}
\end{center}
\caption{Distribution of non-matching probabilities in the cross-match used for the candidate selection (red), a cross-match with artificially shifted positions (blue), and for the known XDINSs (green dashed) that comprise the five \ros\-detected sources on the western Galactic hemisphere and the proposed candidates \ztto\ \citep{2022MNRAS.509.1217R, 2022A&A...666A.148P} and \carINS\ \citep{2015A&A...583A.117P}. }
\label{fig_false_selec_distr}
\end{figure}

\begin{figure}[t]
\begin{center}
\includegraphics[width=\linewidth]{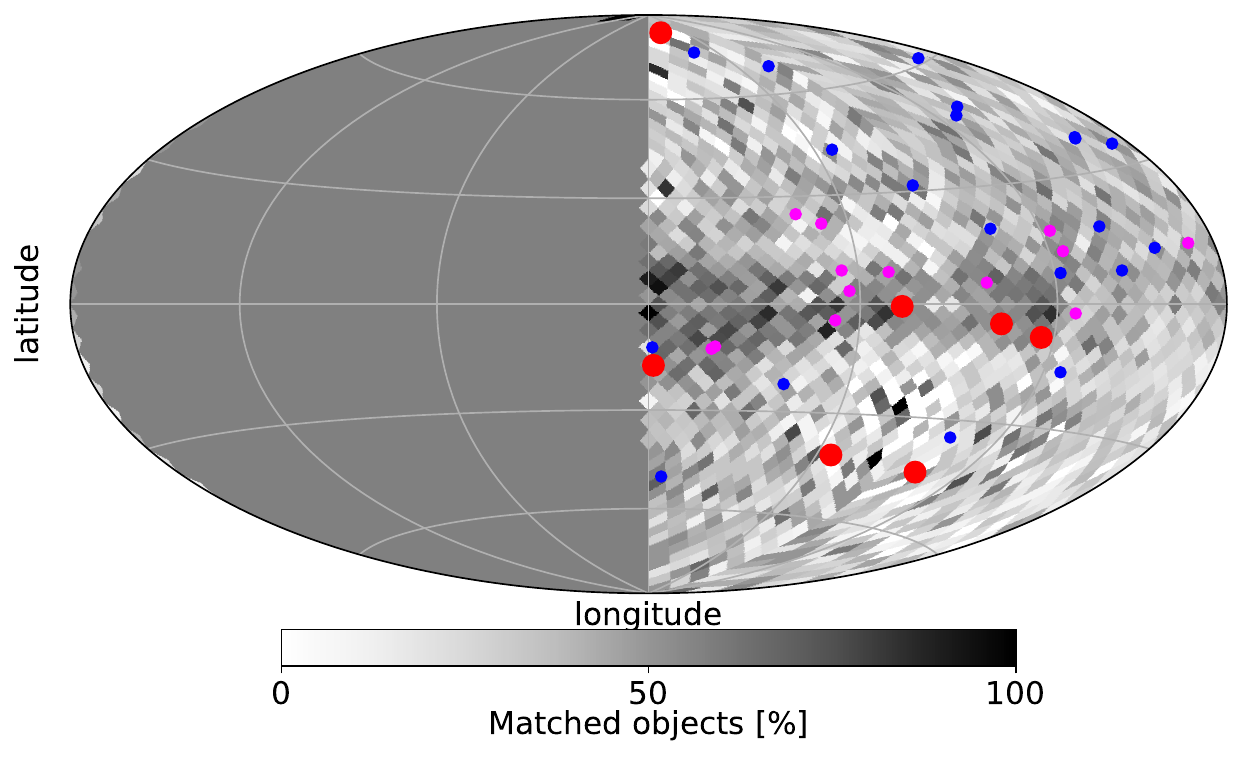}
\end{center}
\caption{Galactic view of the false-deselection rate sky distribution. We show the location of the 13 softer (magenta) and the 20 harder (blue) candidates. The known XDINSs (same sources as in Fig.~\ref{fig_false_selec_distr}) are indicated in red.}
\label{fig_false_sel_skyplot}
\end{figure}

To study the chance of spurious identifications, we assessed the false-deselection rate of INS candidates by describing how likely just by positional coincidence an eRASS source would have matched with an optical or IR counterpart. To estimate this rate, we shifted the \eros\ positions of all 27,670 sources that survived the hardness ratio cut by 1\arcmin\ to the north-east and repeated the ARCHES cross-matching with all optical and IR catalogues. We show the resulting cumulative distribution of non-matching probabilities (blue) in Fig.~\ref{fig_false_selec_distr}. As it can be seen, at the arbitrarily chosen threshold of 50~\% non-matching probability, about 39\% of the shifted sources were erroneously associated with an optical or IR counterpart (as marked by the blue dashed lines), indicating a false-deselection rate of this magnitude. Figure~\ref{fig_false_sel_skyplot} presents the dependence of the false-deselection rate with Galactic position. The Galactic plane, where deep DeCAPS2 imaging is available, can be clearly identified, given the comparatively large number of erroneously matched objects. Quantitatively, at $|b|<15$\degr\ the false-deselection rate is 46~\%. Contrarily at $|b|>15$\degr, the false-deselection fraction decreases to 34~\%.

Lower false-deselection rates can be achieved by decreasing the applied non-matching probability threshold. Doing so will, on the other hand, heavily increase the number of remaining objects. This can be seen from Fig.~\ref{fig_false_selec_distr} where the non-matching probability distribution of the cross-matching based on the original eRASS positions (red line) is also presented. At the current cut of 50\% non-matching probability, we retain 554 sources ($\sim2\%$ of the input sample). If one would attempt to decrease the false-deselection rate to 20\% all-sky, a non-matching probability threshold of 20\% would need to be applied (see blue line in Fig.~\ref{fig_false_selec_distr}). However, this threshold would retain 1542 sources, an increase by a factor $>2.7$.

To further estimate how strongly source confusion may affect the selection of candidates in our sample, we also studied the effectiveness to select already known XDINSs and proposed candidates. In this regard, the selection was very successful since all the five known XDINSs on the western Galactic hemisphere were recovered. Comparing the selection to previous INS searches, we also singled out the INSs \carINS\ \citep[95\% non-matching probability;][]{2015A&A...583A.117P} and \ztto\ \citep[93\% non-matching probability;][]{2022MNRAS.509.1217R, 2022A&A...666A.148P}. That all of these sources survive the cross-match is also shown in Fig.~\ref{fig_false_selec_distr}. Other than the INS \ztto\ confirmed by \citet{2022A&A...666A.148P}, the other three candidates proposed by \citet{2022MNRAS.509.1217R} in the \xmm\ 4XMM-DR10 footprint, \rigi, \rigii, and \rigiii\ have not entered our sample because they were either outside the western Galactic hemisphere or were fainter than the applied flux limit. More recently, \cite{2024ApJ...961...36D} searched the reprocessed version of the \ros\ catalogue \citep{2016A&A...588A.103B} for possible INS candidates. The resulting candidate list includes 55 objects, among them also two known XDINSs. Seventeen out of these 55 sources are located on the western Galactic hemisphere. Adopting a fixed search radius of 1\arcmin, we matched the position of these candidates with the X-ray sources of the stacked eRASS catalogue. We only found possible matches for four of them. Among them, there is the known XDINS \magot\ and three newly proposed candidates: \demi, \demii, and \demiii. The candidates \demii\ and \demiii\ possess \eros\ flux values below the applied flux limit ($\sim 2 \times 10^{-14}$~\fluxcgs, 0.2 -- 2~keV), whereas the source \demi\ is bright, $\sim 2 \times 10^{-12}$~\fluxcgs (0.2 -- 2~keV), but with spectral properties incompatible with our cuts in hardness ratio: $HR1=-0.82(3)$, $HR2=-0.59(8)$, and $HR3=-0.65(17)$. For this source, the angular distance between \ros\ and \eros\ is quite large (49\arcsec), whereas for the other two sources the \ros\ position matched with \eros\ within 20\arcsec. We note that the \eros\ position of \demi\ implies a likely association to the cataclysmic variable (CV) \catvoton. We finally conclude that based on the \eros\ measurements of known XDINSs and recently proposed candidates, we were able to select all of them.

\subsection{X-ray spectral analysis\label{sec_x_ray_fits}}

\subsubsection{Methodology}

We analyzed the spectra of all the 33 selected candidates using the Bayesian X-ray Analysis \citep[BXA;][]{2014A&A...564A.125B} package connected to XSPEC. Specifically, we modeled the candidate spectra with a single blackbody (\texttt{BB}) or a single power law (\texttt{PL}) component in order to study if the continuum of the source is primarily of thermal or non-thermal nature. To account for the interstellar medium absorption in the line-of-sight, we combined the spectral models with the XSPEC \texttt{tbabs} component, adopting the abundance values of \cite{2000ApJ...542..914W}. This model computes the X-ray absorption cross-sections on-the-fly. For the spectral fits, we allowed for generally broad parameter ranges (e.g. $10^{18}$ -- $10^{24}$~cm$^{-2}$ for the hydrogen column density, $\nh$) that contain all the physically acceptable values, and refrained from fixing the model parameters to specific values. We applied uniform priors for parameters that vary within a range spanning less than three decades (e.g. $kT$ or $\Gamma$), while log-uniform priors were adopted for the remaining ones (e.g. $\nh$).

\begin{figure}[t]
\begin{center}
\includegraphics[width=\linewidth]{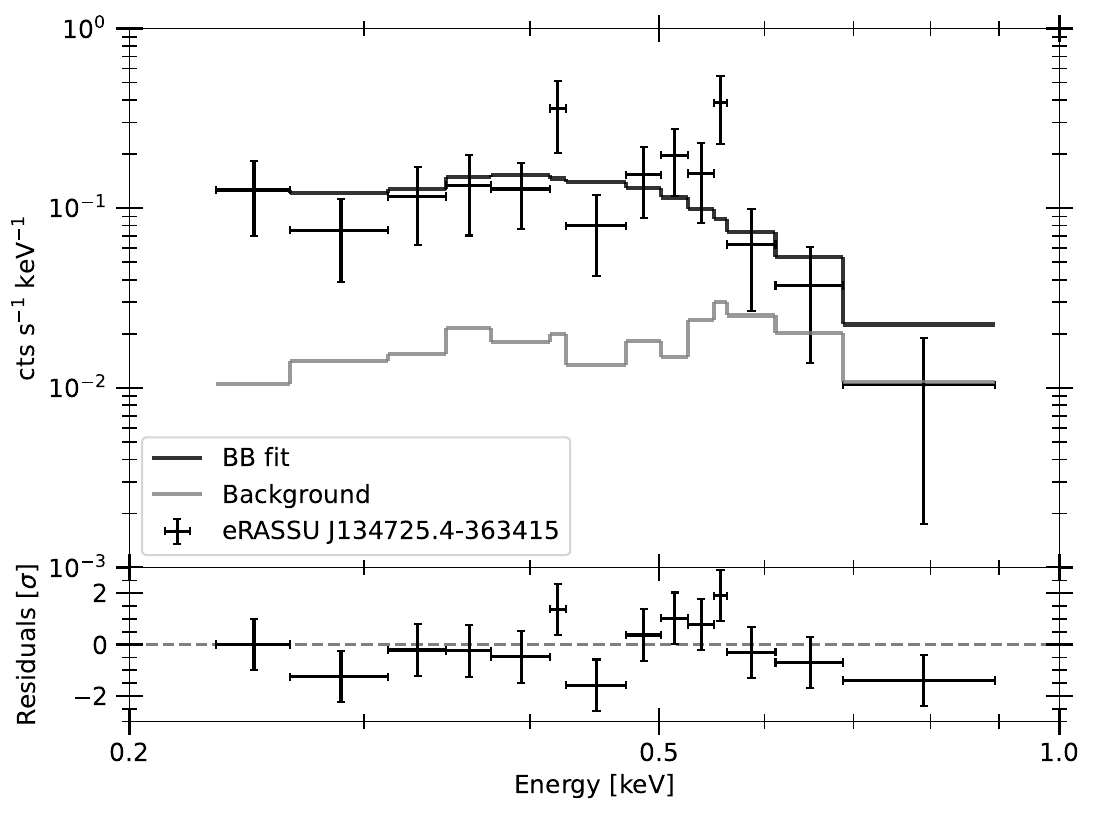}
\end{center}
\caption{Example \eros\ spectrum and \texttt{BB} fit for the source J13472. For presentation purposes, the spectrum was rebinned.}
\label{fig_spec_plots}
\end{figure}

The results of all fits are shown in Table~\ref{tab_specfit}. In order to assess the best model between a \texttt{BB} or \texttt{PL}, we list the resulting Bayesian evidence ($\log(Z)$), the Bayes factor with respect to the \texttt{BB} model ($\log(Z_{BB}/Z)$), and the value for the Akaike Information Criterion (AIC). Generally, models that possess larger Bayesian evidence or smaller AIC values are to be favoured. To judge the significance of these results, we conducted simulations aiming to constrain the false-positive and false-negative rate that the source spectrum is of thermal nature. In the first case, we simulated 250 spectra based on the best-fit \texttt{PL} result to study how often a similarly well or better fitting \texttt{BB} is observed by chance, although the fitted spectrum is of non-thermal origin. For the second case, we simulated 250 spectra based on the best-fit \texttt{BB} solution to estimate how likely the true \texttt{BB} nature could be rejected based on a better fitting \texttt{PL} model. The results of the simulations are displayed in Table~\ref{tab_specfit}. To give an impression on the quality of the \eros\ spectra and the resulting fits, we show the spectrum and subsequent best-fit \texttt{BB} model for J13472 in Fig.~\ref{fig_spec_plots}.

As a consistency test, we included the two XDINS candidates reported in \cite{2023A&A...674A.155K}, \zsfs\ and \otos\ in the spectral analysis. We did not discover any discrepancies in the analysis of either source, indicating that the results presented here are robust. To investigate whether the known light-leak \citep{2021A&A...647A...1P} of two of the seven telescope modules of \eros\ might have affected the analysis, we repeated the fits with spectra containing photons only from the affected modules (TM5 and TM7) and with spectra that should be unaffected, by using events from the remaining 5 modules. These fits, albeit less accurate, are in agreement with the results presented in Table~\ref{tab_specfit} based on TM0. We conclude that counting statistics seems to dominate over systematic effects that might arise due to the light leak for all the observed sources or that the observations were generally not affected by stray light.

Based on the fit results presented in Table~\ref{tab_specfit}, we distinguish two types of sources in our candidate sample: there are soft X-ray emitters ($kT<150$~eV, $\Gamma>5$) and those with slightly harder spectra ($kT\gtrsim200$~eV, $\Gamma\sim 1-4$). For these two groups, we will discuss the results of the spectral analysis independently in the following two subsections.

\subsubsection{Soft X-ray emitters\label{sec_sxe}}

We have identified 13 sources that possess \texttt{BB} spectra with effective temperatures $<150$~eV. Their spectral properties match well with the known XDINSs \citep[$kT\sim40-100$~eV;][]{2007Ap&SS.308..181H}. The corresponding \texttt{PL} fits converge to photon indices with values $>5$, which is too steep for most astronomical sources contained in the eRASS catalogues. In fact, the only sources that may depict such a soft \texttt{PL} like spectrum are narrow-line Seyfert 1 galaxies \cite[NLS1s;][]{2023A&A...669A..37G}. With only one exception, the 13 soft sources possess generally small false-positive rates ($<10$\%), thus implying with high confidence a thermal nature for these sources. For the only exception of J09042, the best-fit \texttt{PL} model is exceptionally steep ($\Gamma\sim 13$), and the $\nh$ is very large. From a physical interpretation of the fit parameters, this makes the \texttt{PL} fit unlikely since there are no astronomical objects that possess such steep \texttt{PL} spectra. This behaviour of the \texttt{PL} fit is often observed for sources where \texttt{BB} models converge to colder effective temperatures. This is also the case for the best-fit \texttt{BB} solution of this source ($kT\sim 100$~eV).

The \texttt{BB} fit results of the 13 sources generally agree with a Galactic nature since the estimates on the $\nh$ are consistent with values lower than the Galactic value. The only exceptions are J09042, J13171 and J17553, where the $\nh$ exceeds the Galactic value by more than $1\sigma$. While this might initially look concerning, these results may not necessarily imply a non-INS nature. At least in the case of J13171, as follow-up observations at X-ray wavelengths have shown \citep[][]{2024A&A...683A.164K}, the large $\nh$ is likely due to a strong absorption feature at soft energies that alters the shape of the spectral continuum. In this case, a simple \texttt{BB} model can only adequately describe the observed spectrum, if it converges towards a stronger ISM absorption. Larger signal-to-noise ratio spectra are needed to lift this degeneracy. We found the obtained \texttt{BB} emission radius values to be generally consistent with an INS nature for all soft candidates since they would agree with the canonical INS radius of 12~km for distances of 1 -- 15~kpc, depending on the source \citep[the true source distance could also be lower since XDINSs were observed to possess smaller emission region sizes, see e.g.][or if the thermal emission originates from a hot spot only]{2002ApJ...572..996D}.

\subsubsection{Harder X-ray emitters}

The remaining 20 candidates possess spectra where \texttt{BB} fits imply effective temperatures $\gtrsim200$~eV. These might be in agreement with the thermal emission seen in younger INSs \citep[e.g. magnetars;][]{2014ApJS..212....6O} or the temperatures of hot spots observed on the surfaces of RPPs \citep[e.g.][]{2020MNRAS.496.5052P}. The \texttt{PL} fits possess photon indices with $1\sigma$ confidence ranges from $\Gamma\sim 1.5-5$. Discerning the exact source nature based on the \eros\ spectra alone will be more difficult for these sources. This is for the reason that the inferred photon index values are not as unphysical as for most of the softer sources in Sect.~\ref{sec_sxe}. In fact, photon indices of 1 -- 3 are typically observed for the magnetospheric emission of RPPs \citep{2009ASSL..357...91B} or in many objects that are driven by accretion processes like active galactic nuclei \citep[AGN; e.g.][]{2010A&A...512A..58I}.

We found false-positive rates ($<10$\%) that statistically imply a thermal nature only for seven candidates (J05215, J07182, J08071, J08535, J10111, J12215, and J21492). At $kT\gtrsim200$~eV the size of the emission regions are generally below 300~m. Under the assumption of an INS nature, this would either imply that the emission originates from a hot spot on the neutron star surface or that the sources are located much farther away than the assumed 1~kpc. The $\nh$ values of the \texttt{BB} fits are also generally in agreement with a Galactic nature. For two sources, temperatures of $\sim1100$~eV were observed. While the $1\sigma$ confidence regions of these temperature values are sufficiently broad to include not as hot INSs, the current limits on the spectral parameters permit a variety of possible source scenarios but do not allow for a conclusive judgement on their exact nature to be made. These sources show hardness ratio values where one would expect cooling INSs to be strongly absorbed, but the spectral fits indicate them to be located relatively nearby due to the possibility of small $\nh$ values with respect to the Galactic value. This might place them at odds with an INS nature and may imply that other spectral models than a \texttt{BB} or \texttt{PL} nature are needed to properly describe their emission.

Small false-negative rates ($<10$\%) that favour the \texttt{PL} fit were obtained for six candidates. These include also the four sources where the Bayes factor is negative and already favours the \texttt{PL}. Interestingly, for half of them (J08070, J11024, J18334) the \texttt{PL} fits converged to low interstellar absorption values ($\nh\sim10^{20}$~cm$^{-2}$) that are below the Galactic limit and thus imply small distances. This could be in agreement with an INS nature (e.g. a RPP). For the remaining three sources (J08525, J09331, J10395) the $\nh$ would also allow for an extragalactic origin.

The remaining seven candidates (J06274, J07263, J08303, J10095, J12182, J13082, and J17541) are equally well fitted with both models. For that reason, the nature of the continuum emission cannot currently be constrained. X-ray follow-up observations will be necessary to better characterise these sources. We note that for J07263 and J08303, the \texttt{BB} fits are quite hot (absolute values of $kT\geq 500$~eV). While the confidence regions may still agree with an INS nature, the \texttt{PL} fits with $\Gamma \sim 2$ are physically more appealing.

\begin{table*}
\caption{eROSITA spectral fit results.
\label{tab_specfit}}
\centering
\scalebox{.71}{
\begin{tabular}{lllllllrrrrr}
\hline\hline
\multicolumn{5}{l}{Soft candidates}\\
\hline
Source & Model & $\nh$ & $N_{\rm H, gal}$\tablefootmark{(a)} & $kT/\Gamma$ & Radius\tablefootmark{(b)} & log(Z) & log(Z$_\mathrm{BB}$/Z) & AIC & Absorbed flux\tablefootmark{(c)} & FPR/FNR\tablefootmark{(d)} &  RATE\_VAR\tablefootmark{(e)}\\
 & & [$10^{20}$~cm$^{-2}$] & [$10^{20}$~cm$^{-2}$] & [eV]/ & [km] & & & & [$10^{-13}$~\fluxcgs] & [\%] & [$\sigma$]\\
\hline
J06571 & \texttt{BB} & $5^{+4}_{-4}$         & 8.26 & $105^{+16}_{-13}$    & $2.0^{+1.3}_{-0.8}$   & -39.08 & 0    & 65.90  & $2.25^{+0.28}_{-0.27}$        &  95.2  & 1.02 \\
J06571 & \texttt{PL} & $16^{+6}_{-4}$        & 8.26 & $6.5^{+0.8}_{-0.8}$  &                       & -42.00 & 2.92 & 70.50  & $2.29^{+0.29}_{-0.26}$        &  0.8   & 1.02 \\
J07230 & \texttt{BB} & $0.6^{+4}_{-0.6}$     & 64.8 & $92^{+12}_{-13}$     & $1.7^{+1.3}_{-0.5}$   & -23.87 & 0    & 37.54  & $1.50^{+0.24}_{-0.23}$        &  99.6  & 1.19 \\
J07230 & \texttt{PL} & $12^{+10}_{-5}$       & 64.8 & $7.1^{+1.8}_{-1.3}$  &                       & -26.00 & 2.13 & 41.37  & $1.36^{+0.2}_{-0.19}$         &  0.4   & 1.19 \\
J08195 & \texttt{BB} & $0.13^{+0.8}_{-0.11}$ & 6.65 & $51^{+4}_{-5}$       & $11.5^{+5}_{-2.6}$    & -24.17 & 0    & 36.63  & $3.8^{+0.5}_{-0.5}$           &  99.6  & 2.14 \\
J08195 & \texttt{PL} & $4.9^{+2.6}_{-2.5}$   & 6.65 & $8.8^{+1.8}_{-1.4}$  &                       & -25.11 & 0.94 & 37.48  & $3.7^{+0.5}_{-0.4}$           &  0.0   & 2.14 \\
J08404 & \texttt{BB} & $0.3^{+2.2}_{-0.4}$   & 5.74 & $80^{+9}_{-9}$       & $2.7^{+1.5}_{-0.8}$   & -18.00 & 0    & 25.66  & $2.1^{+0.4}_{-0.4}$           &  98.8  & 1.41 \\
J08404 & \texttt{PL} & $9^{+5}_{-4}$         & 5.74 & $7.6^{+1.4}_{-1.1}$  &                       & -20.24 & 2.24 & 28.25  & $1.86^{+0.28}_{-0.27}$        &  0.0   & 1.41 \\
J09042 & \texttt{BB} & $90^{+40}_{-40}$      & 27.5 & $102^{+29}_{-19}$    & $9^{+26}_{-7}$        & -29.61 & 0    & 46.06  & $0.96^{+0.16}_{-0.13}$        &  38.4  & 0.36 \\
J09042 & \texttt{PL} & $190^{+40}_{-50}$     & 27.5 & $12.9^{+2.0}_{-2.5}$ &                       & -29.96 & 0.35 & 47.45  & $0.97^{+0.14}_{-0.14}$        &  48.4  & 0.36 \\
J11335 & \texttt{BB} & $7^{+12}_{-8}$        & 14.0 & $132^{+22}_{-25}$    & $0.8^{+1.1}_{-0.4}$   & -26.74 & 0    & 42.33  & $0.99^{+0.12}_{-0.11}$        &  78.8  & 1.77 \\
J11335 & \texttt{PL} & $41^{+18}_{-14}$      & 14.0 & $7.4^{+1.5}_{-1.3}$  &                       & -28.80 & 2.06 & 44.18  & $0.96^{+0.11}_{-0.11}$        &  3.6   & 1.77 \\
J12521 & \texttt{BB} & $4^{+5}_{-4}$         & 37.2 & $138^{+16}_{-14}$    & $0.79^{+0.4}_{-0.23}$ & -41.74 & 0    & 70.15  & $1.48^{+0.14}_{-0.13}$        &  98.4  & 1.61 \\
J12521 & \texttt{PL} & $25^{+8}_{-7}$        & 37.2 & $6.1^{+0.8}_{-0.7}$  &                       & -47.71 & 5.97 & 80.46  & $1.48^{+0.15}_{-0.13}$        &  0.0   & 1.61 \\
J13040 & \texttt{BB} & $22^{+9}_{-8}$        & 20.8 & $96^{+13}_{-12}$     & $3.6^{+4}_{-1.6}$     & -37.68 & 0    & 60.03  & $1.29^{+0.12}_{-0.12}$        &  78.0  & 0.91 \\
J13040 & \texttt{PL} & $57^{+18}_{-13}$      & 20.8 & $9.2^{+1.4}_{-1.2}$  &                       & -40.71 & 3.03 & 66.90  & $1.28^{+0.13}_{-0.12}$        &  0.4   & 0.91 \\
J13171 & \texttt{BB} & $11.7^{+4}_{-2.9}$    & 6.69 & $108^{+7}_{-7}$      & $3.3^{+1.1}_{-0.8}$   & -72.85 & 0    & 128.35 & $4.41^{+0.24}_{-0.22}$        &  87.6  & 1.28 \\
J13171 & \texttt{PL} & $33^{+6}_{-5}$        & 6.69 & $7.4^{+0.5}_{-0.5}$  &                       & -80.92 & 8.07 & 143.84 & $4.42^{+0.24}_{-0.25}$        &  0.0   & 1.28 \\
J13335 & \texttt{BB} & $44^{+10}_{-9}$       & 44.0 & $103^{+10}_{-9}$     & $7.8^{+6}_{-3.0}$     & -85.06 & 0    & 151.46 & $3.62^{+0.19}_{-0.19}$        &  25.6  & 1.51 \\
J13335 & \texttt{PL} & $115^{+18}_{-17}$     & 44.0 & $11.3^{+1.2}_{-1.1}$ &                       & -86.60 & 1.54 & 155.33 & $3.65^{+0.20}_{-0.19}$        &  0.8   & 1.51 \\
J13472 & \texttt{BB} & $0.14^{+0.8}_{-0.12}$ & 4.32 & $89^{+8}_{-8}$       & $1.8^{+0.6}_{-0.4}$   & -34.71 & 0    & 59.06  & $1.55^{+0.18}_{-0.18}$        &  99.2  & 1.77 \\
J13472 & \texttt{PL} & $6.0^{+2.7}_{-2.4}$   & 4.32 & $6.3^{+0.8}_{-0.8}$  &                       & -39.28 & 4.57 & 65.03  & $1.53^{+0.22}_{-0.19}$        &  0.0   & 1.77 \\
J17485 & \texttt{BB} & $4.1^{+1.9}_{-1.8}$   & 12.1 & $115^{+8}_{-7}$      & $2.2^{+0.6}_{-0.5}$   & -56.14 & 0    & 97.08  & $4.6^{+0.4}_{-0.4}$           &  55.6  & 2.81 \\
J17485 & \texttt{PL} & $16.0^{+4}_{-2.7}$    & 12.1 & $6.3^{+0.5}_{-0.4}$  &                       & -60.82 & 4.68 & 104.66 & $4.6^{+0.4}_{-0.4}$           &  0.4   & 2.81 \\
J17553 & \texttt{BB} & $33^{+21}_{-20}$      & 11.1 & $96^{+25}_{-18}$     & $5^{+14}_{-4}$        & -23.37 & 0    & 33.41  & $1.44^{+0.20}_{-0.19}$        &  93.6  & 1.61 \\
J17553 & \texttt{PL} & $100^{+40}_{-40}$     & 11.1 & $11.4^{+2.7}_{-2.4}$ &                       & -25.39 & 2.02 & 38.47  & $1.39^{+0.21}_{-0.17}$        &  0.4   & 1.61 \\
\hline
\multicolumn{5}{l}{Harder candidates}\\
\hline
Source & Model & $\nh$ & $N_{\rm H, gal}$\tablefootmark{(a)} & $kT/\Gamma$ & Radius\tablefootmark{(b)} & log(Z) & log(Z$_\mathrm{BB}$/Z) & AIC & Absorbed flux\tablefootmark{(c)} & FPR/FNR\tablefootmark{(d)} &  RATE\_VAR\tablefootmark{(e)}\\
 & & [$10^{20}$~cm$^{-2}$] & [$10^{20}$~cm$^{-2}$] & [eV]/ & [km] & & & & [$10^{-13}$~\fluxcgs] & [\%] & [$\sigma$]\\
\hline
J05215 & \texttt{BB} & $0.07^{+0.28}_{-0.06}$ & 2.02 & $206^{+12}_{-11}$      & $0.25^{+0.03}_{-0.029}$   & -61.51 & 0     & 111.66 & $1.17^{+0.09}_{-0.10}$    &  99.2  & 2.08 \\
J05215 & \texttt{PL} & $2.2^{+2.0}_{-1.8}$    & 2.02 & $3.29^{+0.28}_{-0.26}$ &                           & -70.33 & 8.82  & 125.74 & $1.95^{+0.14}_{-0.15}$    &  0.0   & 2.08 \\
J06274 & \texttt{BB} & $0.13^{+0.8}_{-0.11}$  & 7.71 & $198^{+23}_{-18}$      & $0.27^{+0.07}_{-0.06}$    & -29.80 & 0     & 51.49  & $1.14^{+0.19}_{-0.16}$    &  22.8  & 2.54 \\
J06274 & \texttt{PL} & $2.1^{+7}_{-2.1}$      & 7.71 & $3.3^{+0.8}_{-0.5}$    &                           & -30.19 & 0.39  & 50.04  & $1.94^{+0.28}_{-0.24}$    &  27.6  & 2.54 \\
J07182 & \texttt{BB} & $2.6^{+28}_{-2.6}$     & 9.61 & $1100^{+600}_{-600}$   & $0.021^{+0.021}_{-0.007}$ & -11.96 & 0     & 18.89  & $6.6^{+1.6}_{-1.4}$       &  100.0 & 1.18 \\
J07182 & \texttt{PL} & $110^{+100}_{-60}$     & 9.61 & $4.1^{+2.9}_{-1.8}$    &                           & -13.75 & 1.79  & 18.67  & $2.1^{+0.5}_{-0.5}$       &  0.0   & 1.18 \\
J07263 & \texttt{BB} & $0.27^{+5}_{-0.25}$    & 14.1 & $500^{+500}_{-140}$    & $0.056^{+0.04}_{-0.025}$  & -23.98 & 0     & 41.47  & $1.7^{+0.4}_{-0.4}$       &  14.0  & 1.09 \\
J07263 & \texttt{PL} & $0.7^{+9}_{-0.7}$      & 14.1 & $2.0^{+0.7}_{-0.5}$    &                           & -24.37 & 0.39  & 41.32  & $6.5^{+1.2}_{-1.3}$       &  70.0  & 1.09 \\
J08070 & \texttt{BB} & $0.13^{+1.0}_{-0.11}$  & 4.07 & $260^{+50}_{-40}$      & $0.16^{+0.06}_{-0.05}$    & -33.70 & 0     & 60.85  & $1.35^{+0.25}_{-0.22}$    &  8.4   & 2.34 \\
J08070 & \texttt{PL} & $0.18^{+1.5}_{-0.16}$  & 4.07 & $2.7^{+0.4}_{-0.4}$    &                           & -34.18 & 0.48  & 59.04  & $3.3^{+0.6}_{-0.5}$       &  48.0  & 2.34 \\
J08071 & \texttt{BB} & $0.25^{+2.5}_{-0.23}$  & 9.66 & $204^{+40}_{-24}$      & $0.26^{+0.09}_{-0.07}$    & -29.09 & 0     & 51.51  & $1.13^{+0.19}_{-0.17}$    &  100.0 & 2.57 \\
J08071 & \texttt{PL} & $5^{+40}_{-5}$         & 9.66 & $3.3^{+2.2}_{-0.7}$    &                           & -32.81 & 3.72  & 56.36  & $1.94^{+0.4}_{-0.28}$     &  0.0   & 2.57 \\
J08303 & \texttt{BB} & $0.27^{+2.3}_{-0.24}$  & 4.49 & $610^{+700}_{-240}$    & $0.040^{+0.04}_{-0.020}$  & -17.79 & 0     & 30.79  & $2.4^{+0.6}_{-0.6}$       &  47.2  & 0.96 \\
J08303 & \texttt{PL} & $1.0^{+19}_{-1.0}$     & 4.49 & $1.9^{+0.9}_{-0.6}$    &                           & -18.56 & 0.77  & 30.15  & $7.3^{+1.6}_{-1.6}$       &  55.2  & 0.96 \\
J08525 & \texttt{BB} & $0.15^{+1.2}_{-0.14}$  & 2.88 & $240^{+50}_{-40}$      & $0.18^{+0.08}_{-0.06}$    & -24.57 & 0     & 42.94  & $1.20^{+0.23}_{-0.21}$    &  0.4   & 3.16 \\
J08525 & \texttt{PL} & $0.3^{+2.8}_{-0.28}$   & 2.88 & $3.0^{+0.5}_{-0.4}$    &                           & -23.06 & -1.51 & 37.88  & $2.6^{+0.5}_{-0.5}$       &  89.2  & 3.16 \\
J08535 & \texttt{BB} & $1.3^{+23}_{-1.3}$     & 2.78 & $1100^{+600}_{-600}$   & $0.023^{+0.023}_{-0.007}$ & -17.07 & 0     & 27.90  & $7.8^{+1.6}_{-1.4}$       &  99.2  & 1.31 \\
J08535 & \texttt{PL} & $60^{+110}_{-60}$      & 2.78 & $2.4^{+3.0}_{-1.9}$    &                           & -18.50 & 1.43  & 27.49  & $7.9^{+1.6}_{-1.5}$       &  1.6   & 1.31 \\
J09331 & \texttt{BB} & $0.12^{+0.9}_{-0.11}$  & 3.83 & $227^{+22}_{-19}$      & $0.28^{+0.07}_{-0.05}$    & -32.22 & 0     & 56.37  & $2.25^{+0.29}_{-0.26}$    &  6.0   & 1.98 \\
J09331 & \texttt{PL} & $5^{+8}_{-5}$          & 3.83 & $3.4^{+0.7}_{-0.5}$    &                           & -32.36 & 0.14  & 52.30  & $3.5^{+0.5}_{-0.4}$       &  33.2  & 1.98 \\
J10095 & \texttt{BB} & $0.19^{+1.5}_{-0.17}$  & 3.62 & $210^{+50}_{-40}$      & $0.21^{+0.10}_{-0.07}$    & -15.59 & 0     & 25.61  & $0.89^{+0.21}_{-0.18}$    &  43.2  & 1.43 \\
J10095 & \texttt{PL} & $1.4^{+9}_{-1.4}$      & 3.62 & $3.3^{+0.9}_{-0.6}$    &                           & -16.29 & 0.7   & 24.38  & $1.5^{+0.4}_{-0.4}$       &  23.6  & 1.43 \\
J10111 & \texttt{BB} & $0.3^{+2.4}_{-0.27}$   & 2.28 & $270^{+70}_{-50}$      & $0.15^{+0.08}_{-0.05}$    & -19.96 & 0     & 34.91  & $1.28^{+0.26}_{-0.26}$    &  95.2  & 2.23 \\
J10111 & \texttt{PL} & $5^{+13}_{-5}$         & 2.28 & $2.7^{+1.2}_{-0.7}$    &                           & -21.58 & 1.62  & 33.68  & $3.5^{+0.7}_{-0.7}$       &  4.0   & 2.23 \\
J10395 & \texttt{BB} & $0.18^{+1.3}_{-0.16}$  & 1.58 & $380^{+260}_{-90}$     & $0.07^{+0.05}_{-0.04}$    & -22.79 & 0     & 39.20  & $1.00^{+0.24}_{-0.21}$    &  0.4   & 1.46 \\
J10395 & \texttt{PL} & $0.25^{+2.5}_{-0.22}$  & 1.58 & $2.5^{+0.6}_{-0.5}$    &                           & -21.05 & -1.74 & 34.88  & $2.5^{+0.7}_{-0.6}$       &  96.4  & 1.46 \\
J11024 & \texttt{BB} & $0.07^{+0.4}_{-0.05}$  & 4.66 & $189^{+16}_{-16}$      & $0.39^{+0.08}_{-0.07}$    & -37.72 & 0     & 66.32  & $2.04^{+0.24}_{-0.23}$    &  8.4   & 1.03 \\
J11024 & \texttt{PL} & $0.7^{+2.4}_{-0.7}$    & 4.66 & $3.32^{+0.4}_{-0.27}$  &                           & -37.64 & -0.08 & 63.98  & $3.4^{+0.5}_{-0.4}$       &  10.4  & 1.03 \\
J12182 & \texttt{BB} & $0.18^{+1.5}_{-0.16}$  & 1.36 & $310^{+90}_{-60}$      & $0.10^{+0.05}_{-0.04}$    & -23.67 & 0     & 42.58  & $0.92^{+0.20}_{-0.19}$    &  15.2  & 1.11 \\
J12182 & \texttt{PL} & $0.31^{+4}_{-0.29}$    & 1.36 & $2.5^{+0.6}_{-0.5}$    &                           & -24.16 & 0.49  & 40.25  & $2.9^{+0.7}_{-0.6}$       &  60.4  & 1.11 \\
J12215 & \texttt{BB} & $0.17^{+1.2}_{-0.15}$  & 3.58 & $207^{+21}_{-19}$      & $0.27^{+0.07}_{-0.06}$    & -30.01 & 0     & 51.60  & $1.36^{+0.18}_{-0.17}$    &  74.0  & 1.36 \\
J12215 & \texttt{PL} & $4^{+12}_{-4.0}$       & 3.58 & $3.6^{+0.9}_{-0.5}$    &                           & -31.90 & 1.89  & 52.84  & $2.03^{+0.26}_{-0.24}$    &  3.6   & 1.36 \\
J13082 & \texttt{BB} & $0.14^{+0.9}_{-0.12}$  & 1.87 & $251^{+29}_{-25}$      & $0.17^{+0.05}_{-0.04}$    & -31.62 & 0     & 56.01  & $1.18^{+0.17}_{-0.16}$    &  42.4  & 2.59 \\
J13082 & \texttt{PL} & $1.1^{+6}_{-1.1}$      & 1.87 & $2.9^{+0.5}_{-0.4}$    &                           & -32.79 & 1.17  & 55.03  & $2.4^{+0.4}_{-0.4}$       &  12.4  & 2.59 \\
J17541 & \texttt{BB} & $0.22^{+1.6}_{-0.19}$  & 5.75 & $204^{+25}_{-22}$      & $0.23^{+0.07}_{-0.06}$    & -27.23 & 0     & 47.33  & $0.92^{+0.15}_{-0.13}$    &  22.0  & 1.99 \\
J17541 & \texttt{PL} & $6^{+13}_{-6}$         & 5.75 & $3.6^{+1.1}_{-0.6}$    &                           & -27.72 & 0.49  & 44.97  & $1.38^{+0.20}_{-0.21}$    &  32.4  & 1.99 \\
J18334 & \texttt{BB} & $0.17^{+1.4}_{-0.15}$  & 8.80 & $420^{+270}_{-110}$    & $0.07^{+0.05}_{-0.04}$    & -27.08 & 0     & 47.55  & $1.45^{+0.3}_{-0.29}$     &  0.0   & 1.13 \\
J18334 & \texttt{PL} & $0.25^{+2.3}_{-0.23}$  & 8.80 & $2.6^{+0.5}_{-0.5}$    &                           & -24.39 & -2.69 & 40.85  & $3.3^{+0.7}_{-0.6}$       &  100.0 & 1.13 \\
J21492 & \texttt{BB} & $0.23^{+2.4}_{-0.21}$  & 1.39 & $330^{+70}_{-60}$      & $0.089^{+0.04}_{-0.025}$  & -19.59 & 0     & 34.70  & $0.97^{+0.21}_{-0.18}$    &  83.6  & 1.51 \\
J21492 & \texttt{PL} & $0.7^{+13}_{-0.7}$     & 1.39 & $2.5^{+0.7}_{-0.5}$    &                           & -21.11 & 1.52  & 34.34  & $2.8^{+0.7}_{-0.6}$       &  6.8   & 1.51 \\
\hline
\end{tabular}}
\tablefoot{For the absolute parameters, the median value of the sample parameter distribution is shown. The errors mark the 1$\sigma$ confidence region. 400 live points were adopted to sample the parameter space.
\tablefoottext{a}{Galactic hydrogen column density ($\nh$) at the location of the candidates \citep{2016A&A...594A.116H}.}
\tablefoottext{b}{The blackbody emission radius at infinity is computed assuming a 1~kpc source distance.}
\tablefoottext{c}{The absorbed model flux is provided in the 0.2 -- 10~keV band.}
\tablefoottext{d}{False-positive (FPR) and false-negative (FNR) rates. In rows presenting the \texttt{BB} best-fit results, the false-negative rate is given (since this rate is based on simulations with the best-fit \texttt{BB} model), whereas rows presenting \texttt{PL} results provide the false-positive rate (which is estimated from simulations using the best-fit \texttt{PL} model).}
\tablefoottext{e}{Variability parameter computed from the variation in observed source count rate (0.2 -- 2~keV) via $\mathrm{RATE\_VAR}=\underset{k,l \in [1,n]}{\max} \frac{|R_k-R_l|}{\sqrt{\sigma_k^2+\sigma_l^2}}$.}
}
\end{table*}


\subsection{Short- and long-term variability}

The fact that \eros\ repeatedly observed the sources allowed us to search for the presence of long-term variability. This is interesting to investigate since XDINSs are expected to be rather stable X-ray emitters, while other neutron star types (e.g. magnetars) or contaminant sources, such as X-ray binaries (XRBs), CVs, or AGN, would be expected to display significant variation. We computed for each source and each individual survey, the observed count rate in the 0.2 -- 2~keV band, and between all surveys, we calculated the maximum observed rate deviation,

\begin{displaymath}
\mathrm{RATE\_VAR}=\underset{k,l \in [1,n]}{\max} \frac{|R_k-R_l|}{\sqrt{\sigma_k^2+\sigma_l^2}},
\end{displaymath}

by analogy with the definition of the FLUXVAR parameter in the \xmm\ stacked catalogues \citep[e.g.][]{2020A&A...641A.137T}. The results are presented in Table~\ref{tab_specfit}. Only J08525 was observed to vary by more than $3\sigma$, which might be regarded significant. Variability below the $3\sigma$ threshold is also observable for some of the XDINSs on the western Galactic hemisphere. As \cite{2022A&A...666A.148P} argue, these variations in count rate may be attributed to low count statistics. Based on the rate values presented in Appendix B of \cite{2022A&A...666A.148P}, we exemplarily estimated a "RATE\_VAR" of 2.5$\sigma$ for \magzs. However, as is the case for all the known XDINSs, \magzs\ is observed to be intrinsically stable on long time-scales. It was only observed to show changes in spectral state \citep{2012MNRAS.423.1194H}. We thus conclude that most of the selected candidates conform with a stable nature over the course of the \eros\ observations.

\begin{figure}[t]
\begin{center}
\includegraphics[width=\linewidth]{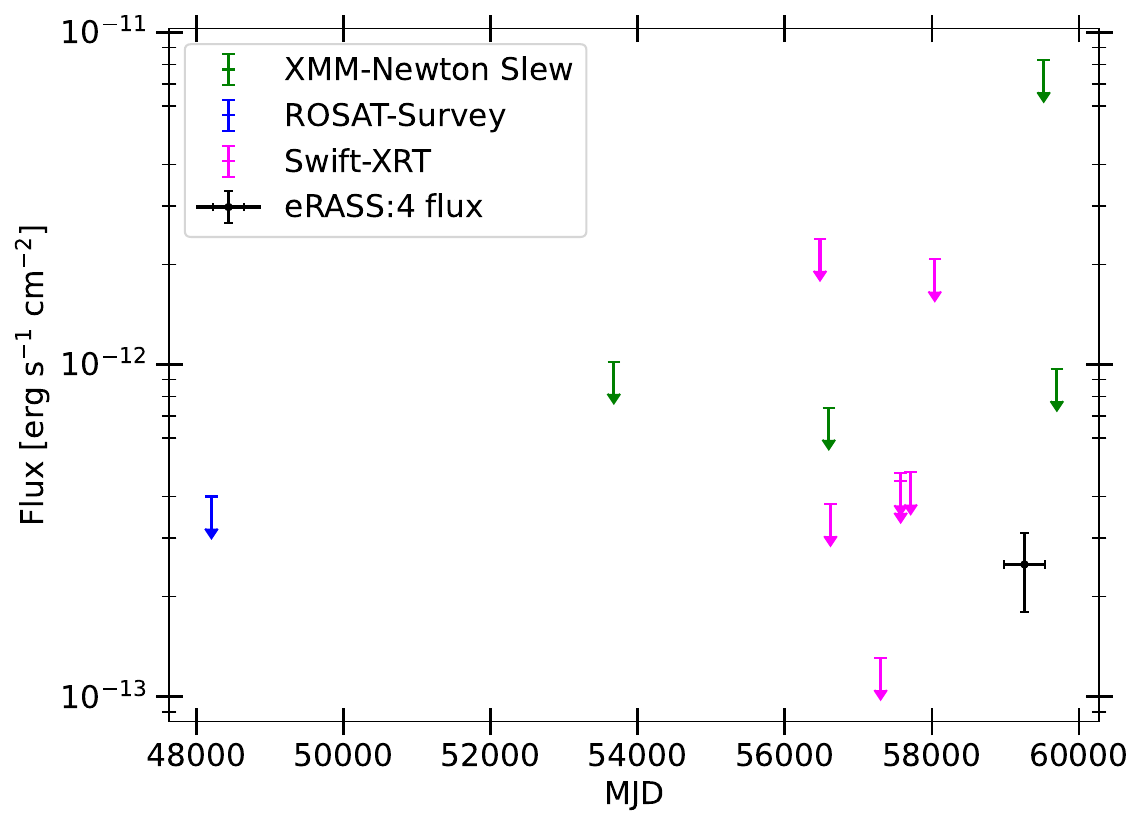}
\end{center}
\caption{Long-term light curve for the candidate J10395. Upper limits from the XMM-Newton Slew Survey (green), the ROSAT All-Sky survey (blue), and \sxrt\ (magenta) are shown. The stacked \eros\ flux is indicated in black, with the x-axis error depicting the time between the first and last \eros\ observation.}
\label{fig_var_lc}
\end{figure}

Checking for variability beyond the \eros\ surveys is made possible by using the HILIGT tool \citep{2022A&C....3800531S,2022A&C....3800529K} to create long-term light curves. These display either detected flux values in past X-ray observations near the \eros\ position or, in the absence of a detection, a $3\sigma$ upper limit. The light curves include detections and upper limit values from \xmm\ pointed observations \citep[e.g.][]{2020A&A...641A.136W}, the XMM-Newton Slew Survey \citep{2008A&A...480..611S}, \sxrt\ \citep{2005SSRv..120..165B}, and \ros\ PSPC and HRI pointed observations, as well as from the PSPC All-Sky Survey \citep{1987SPIE..733..519P,2016A&A...588A.103B}. The HILIGT tool bases its computation on a pre-defined grid of parameter values. As such, for each candidate we individually determined the grid point that is closest to the best-fit solution presented in Table~\ref{tab_specfit}. The resulting light curves were then checked for significant variability. We found 11 of our candidates to be detected by \ros\ and three of them to be also detected in the XMM-Newton Slew Survey (J08195, J13171, J17485). In all cases, the archival X-ray flux values were in agreement with \eros. The presence of variability was only indirectly observed for three undetected sources (J09331, J10395, and J11024), where archival upper limits were observed that reached below the \eros\ flux values. Such a light curve is presented for J10395 in Fig.~\ref{fig_var_lc}. Interestingly, all three sources belong to the harder sample and a non-thermal spectral continuum seems more likely. While J08525 might be variable on shorter timescales, its HIGLIGT light curve does not indicate long term variability since the source was never detected in the past and the upper limits from \ros\ and \xmm\ are generally above the \eros\ flux level.

The overall small number of detected counts that is spread over 4 -- 5 epochs prevents a meaningful timing analysis \citep[see also Sect. 3.2 in][for more information]{2023A&A...674A.155K}. Follow-up observations will be necessary to identify the neutron star spin period, which is essential for deeper insights into the nature of the candidates.

\subsection{X-ray-to-optical flux ratios\label{sec_fxfopt}}

To study if a non-INS nature can be excluded based on archival optical imaging alone, we used the computed magnitude limits to estimate the X-ray-to-optical flux ratio. To allow for comparison of the limiting magnitudes calculated in different passbands (either \gaia\  $g$-band or the $r$-band used in Legacy Survey, Pan-STARRS or DeCAPS2), we assumed an optical spectrum that is flat in frequency space and normalised to the observed flux density as computed from the observed magnitude limits. The X-ray-to-optical flux ratio can then be computed via $\log(f_\mathrm{X}/f_\mathrm{opt}) = \log(\frac{F_\mathrm{X}}{\mathrm{erg}\,\mathrm{s}^{-1}\mathrm{cm}^{-2}}) +\frac{m}{2.5}+5.39$, with $F_\mathrm{X}$ being the X-ray flux, $m$ the magnitude limit and $5.39$ a constant to relate the X-ray-to-optical flux ratio values to the $V$-band. We applied this relation to all the candidates and present the individual results in Table~\ref{tab_cands_prop}. For the X-ray flux the best-fit \texttt{BB}/\texttt{PL} model value was applied (depending on the estimated false-positive and false-negative rates as described in Sect.~\ref{sec_x_ray_fits}; the \texttt{BB} flux was used for the eight candidates that are equally well fit with a \texttt{BB} or \texttt{PL} continuum).

\begin{figure}[t]
\begin{center}
\includegraphics[width=\linewidth]{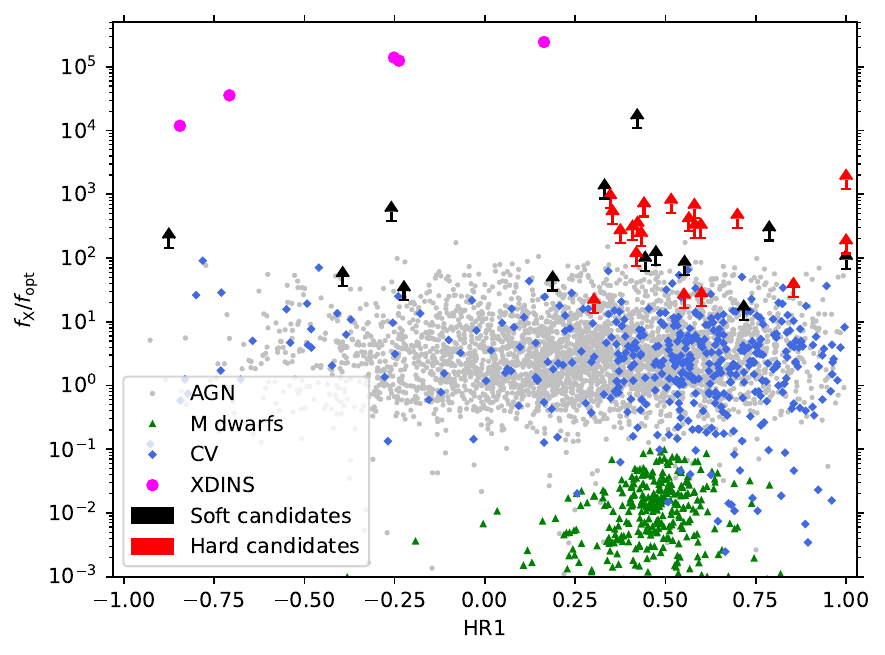}
\end{center}
\caption{X-ray-to-optical flux ratio limits for all the soft (black) and harder (red) candidates and some of the most prevalent contaminants. The limits for \zsfs\ and \otos\ were taken from \cite{2023A&A...674A.155K} and \cite{2024A&A...683A.164K}. The indicated contaminants are the same as were used in Fig.~7 of \cite{2023A&A...674A.155K}.}
\label{fig_fxfopt}
\end{figure}

In Fig.~\ref{fig_fxfopt} we compare the resulting X-ray-to-optical flux ratio values to those of the most abundant contaminants in our search. We found that an M-star nature can be excluded for all of the candidates; however, the majority of candidates (19 sources) are still located in the upper part of the AGN and CV cloud. Thus, based on X-ray-to-optical flux ratio alone, excluding these non-INS natures is difficult. We want to note that the remaining 14 candidates possess X-ray-to-optical flux ratios in excess of 200. Of these, three sources belong to the soft sample of candidates (J06571, J13171, and J13472). Of the remaining 11 sources, seven possess spectra that were equally or better fitted by a \texttt{BB} model (J05215, J08303, J08535, J10111, J12215, J13082, and J21492). While X-ray-to-optical flux ratios above 200 are comparably large and seem to be less common in AGN and CVs, it needs to be considered that also other compact object types, like XRBs or super-soft sources (SSS) may reach similarly large X-ray-to-optical flux ratios. We conclude that deeper optical follow-up observations are required to confirm the INS nature for most candidates, as they have exemplarily already been conducted for J13171 \citep{2024A&A...683A.164K}. Consequently, this source possesses the largest X-ray-to-optical flux ratio in Fig.~\ref{fig_fxfopt}, which confirms the INS nature.

\section{Discussion\label{sec_disc}}

In this work, we have reported the results of a search for thermally emitting INS candidates in the data of the eRASS. The selection criteria, where we screened for soft X-ray emitting sources lacking counterparts at optical and IR wavelengths, have resulted in a candidate sample of 33 sources. We observed the seven established XDINSs to possess thermal spectra with \texttt{BB} temperatures of $\sim 40-100$~eV \citep{2009ASSL..357..141T}. We found 13 promising candidates (see upper section of Table~\ref{tab_specfit}) displaying similarly soft spectra, and if confirmed, they could be in a similar evolutionary state. The spectral analysis of eRASS data remarkably favours (in all but one case) a thermal \texttt{BB} continuum with effective temperatures between $50$~eV and $150$~eV. At the same time, a \texttt{PL} nature may also be seen as unlikely based on the observed \texttt{PL} photon indexes that are oftentimes in excess of five to six for these sources, a limit known to be reached only by a few extreme NLS1s \citep{2023A&A...669A..37G}. We note, however, that present X-ray-to-optical flux ratio limits are still too shallow for most of them to secure an INS nature (see Fig.~\ref{fig_fxfopt}).

While some INS types may exhibit significant variability, most thermally emitting INSs (including XDINS, RPPs and most central compact objects in supernova remnants) are observed to be stable X-ray emitters over long timescales. We found no significant evidence of flux variability in the eRASS data of virtually any soft candidate that could favour an alternative nature, for instance, an SSS \citep[e.g.][reporting the detection of four SSS in \eros]{2022A&A...657A..26M}. Interestingly, all soft candidates are located within $|b| <25$\degr\ from the Galactic plane, where one would expect to find most INSs \citep[e.g.][]{2008A&A...482..617P,2017AN....338..213P}.

Interesting from a population point of view is the discovery of hotter and potentially younger XDINSs, which would allow testing of evolutionary links with other families of INSs. These could be hidden amidst the sample of 20 hotter candidates we selected from eRASS. Of them, seven sources are better fit by a \texttt{BB} model of temperatures of $kT\gtrsim200$~eV with respect to a \texttt{PL} with photon index 2.5 -- 4. The small implied \texttt{BB} emission regions, at an assumed 1~kpc distance, may be indicative of hot spots on the neutron star surfaces. At larger distances, $d\geq 5$~kpc, the X-ray flux is consistent with emission from most of the neutron star surface; such large distances are, however, in disagreement with the generally low column density derived for most of them from the spectra.

The sample of harder candidates in the lower section of Table~\ref{tab_specfit} includes all six objects for which the false-negative rates imply a non-thermal continuum as well as most other candidates for which both a \texttt{BB} and \texttt{PL} model fit the data equally well. The result of a better-fitting \texttt{PL} does not necessarily imply a non-INS nature \citep[see e.g.][]{2009ASSL..357...91B}.

We found that many sources in the hard sample are located at higher Galactic latitudes, implying nearby low-luminosity sources or an extragalactic nature for them. The safest route to distinguish INSs from AGN or an accreting binary system is via the X-ray-to-optical flux ratio. Eleven sources in the hard sample already possess an X-ray-to-optical flux ratio limit above 200. However, also at this level, a CV or AGN nature cannot be fully ruled out. Among the eleven high X-ray-to-optical flux ratio sources are seven that are equally or better fitted with a \texttt{BB} model (J05215, J08303, J08535, J10111, J12215, J13082, and J21492). The remaining four high X-ray-to-optical flux ratio sources are better described by a \texttt{PL} nature. Studying the variability of these four sources, the absence of any significant variation in J08070 and an $\nh$ value that implies a Galactic nature makes it a prime candidate to be an RPP. The remaining three sources were observed to possibly be variable during the two-year period of eRASS data analysed here (J08525) and at longer timescales, considering upper limits by other X-ray instruments (J10395, J11024). This behaviour may be more in agreement with AGN or CVs. It is interesting to note that all candidates that we found to show hints of variability seem to correspond better with a non-thermal nature. The nine sources in the hard sample with low X-ray-to-optical flux ratios require additional follow-up observations to judge their nature.

\begin{figure}[t]
\begin{center}
\includegraphics[width=\linewidth]{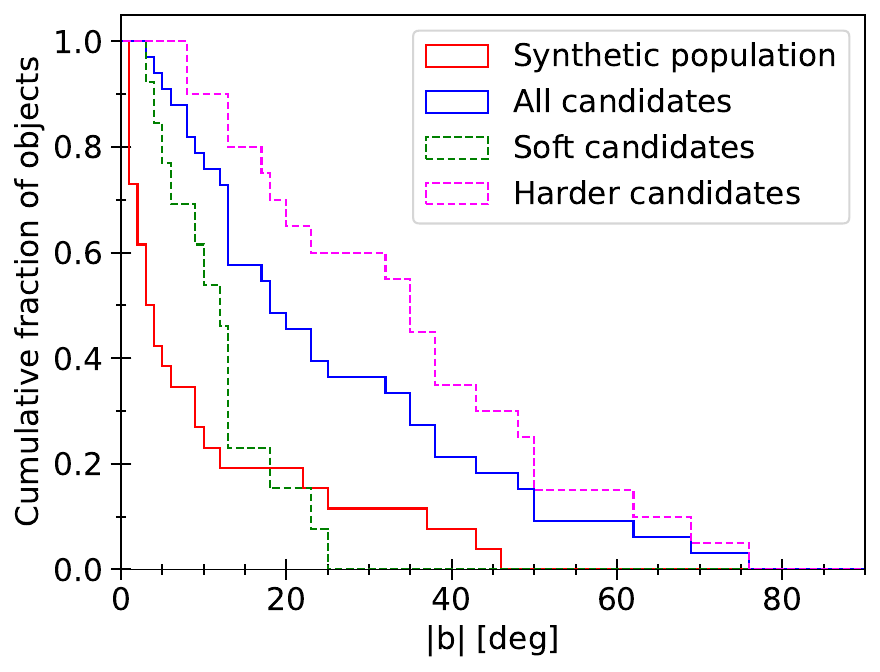}
\end{center}
\caption{Distribution of Galactic latitude ($|b|$), the synthetic population simulated in \citet[][red full]{2017AN....338..213P}, and all identified candidates (blue full). To facilitate more detailed studies, we also show the distribution of the candidates identified as soft emitters ($kT<150$~eV; green dashed) and as harder emitters (magenta dashed).
\label{fig_gal_b_distr}}
\end{figure}


As SRG/\eros\ is primarily a surveying mission, systematic positional inaccuracies or shifts might be expected. Such was the case for the XIDNS candidate J13171. The \xmm\ follow-up observation of the source revealed its position to be consistent only at the $2\sigma$ level with \eros\ \citep{2024A&A...683A.164K}. A similar situation is likely not a concern for most of our candidates, as the angular distances to the nearest optical or IR object are significantly larger (see Table~\ref{tab_cands_prop}). On the other hand, source confusion is critical for the candidates located in crowded areas (like J12521). As such, follow-up observations at X-ray energies are necessary to obtain accurate astrometry while searching for possible optical or IR counterparts.

A promising result of our selection procedure is that we recover all known XDINSs and the recently proposed candidates in the searched \eros\ footprint. As these are generally found at brighter X-ray fluxes, they consequently possess more accurate positions than the majority of \eros\ sources. Thus, the selection accuracy may significantly decrease towards the flux limit of our search ($10^{-13}$~\fluxcgs, 0.2 -- 2~keV). With the cut in hardness-ratio space being fairly inclusive, we expected most XDINSs to be erroneously deselected in the probabilistic cross-matching stage, which is rather strict. As our estimate of the false-deselection rate has shown, more than a third of the population may be erroneously excluded; at low Galactic latitudes, this fraction increases to 50\%.

In \citet{2017AN....338..213P}, neutron stars are born from progenitor massive stars distributed in the spiral arms of the Milky Way, at a scale height of 50~pc in the plane. Consequently, at a timescale relevant for cooling, most neutron stars will be located still relatively close to their birth place, where we expect our search to be most inefficient due to source confusion. In Fig.~\ref{fig_gal_b_distr} we show the distribution in Galactic latitude of the simulated INS population and that of our sample of XDINS candidates. As it can be seen, the distribution of the synthetic neutron stars is considerably more skewed towards low Galactic latitudes than that of our eRASS candidate sample. However, when splitting the candidates into the soft and hard samples, the results look more promising. In fact, all of the soft eRASS candidates are located at low Galactic latitudes ($|b|<25\degr$), which is in agreement with the expectation. The hard sample, on the other hand,  spans a much wider latitude range, which might suggest an alternative nature.

In \citet{2008A&A...482..617P}, the spatial distribution of fainter XDINSs is instead directly tied to those of more remote OB associations of the Milky Way beyond the Gould Belt. Based on this, the authors aim to identify the sky regions that are most promising for XDINS searches. Since, the considered OB associations are located near the Galactic plane, most new XDINSs are likewise expected to be situated at $|b|<30\degr$. In our eRASS sample of XDINS candidates, more than 60\% of the sources would be found at these latitudes, including all candidates of the soft sub-sample.

Within the limits enforced by small number statistics, the eRASS forecast in \citet{2017AN....338..213P} may be used to estimate the number of XDINSs that are in the candidate sample. Given that we have access to only half the sky, one may expect the eRASS input catalogue to contain 9 -- 12 XDINSs. This is derived from the eRASS forecast, where between the western and eastern Galactic hemispheres a ratio of 2:3 XDINSs (above a flux limit of $\sim 2.2 \times 10^{-13}$~\fluxcgs) was found and a total number of $26\pm4$ all-sky was predicted \citep{2017AN....338..213P}. Folding these numbers with the determined false-deselection rates, one may expect that 2 -- 6 XDINSs were matched away. XDINSs are expected to be mostly located near the Galactic plane \citep{2008A&A...482..617P}; thus, it may be more likely that the true numbers are at the larger end of this distribution. Consequently, one may expect that the candidate sample will contain between one and three new XDINSs.

Needless to say, population synthesis models, though a fundamental tool in astrophysics to compare observational results (e.g. from surveys) with theoretical expectations, are subject to several theoretical and observational uncertainties. If a significant number of objects are known, it allows for the testing of many of the population's physical (and theoretically unknown) properties. Very often, however, the number of poorly constrained physical parameters is so large that very different models might be consistent with a given set of observational data. Conversely, the large number of hypotheses and simplifying assumptions tend to leave room for caution about the absolute predictive power of a given model. In this regard, the characterisation of XDINS candidates from the eRASS survey is of utmost importance \citep[see e.g. the case of old INSs expected to be detected through accretion from the interstellar medium by \ros;][for a review]{2000PASP..112..297T}.

Finally, to improve completeness, the goal must be to expand the search towards lower non-matching thresholds. As can be seen in Fig.~\ref{fig_false_selec_distr}, decreasing the probability cut from 50\% to include all sources above 20\% non-matching probability would halve the false-deselection rate (from about 40\% to 20\% all-sky). However, this will increase the number of candidates by a factor $>2.7$. Consequently, the time needed to conduct follow-up observations, which will allow the candidate sample to be fully characterised, will also grow accordingly. As such, searches for XDINSs at low non-matching probabilities may be expected to become more inefficient. To counter this inefficiency, a possible way forward could be to mainly concentrate on the regions that were identified in the past to be most promising to contain new XDINSs \citep{2008A&A...482..617P}.

\section{Conclusions and outlook\label{sec_concl}}

For the first time since the \ros\ era, flux-limited all-sky searches for new thermally emitting INSs covering considerable areas of the sky can be pursued. Our investigation based on the two first years of the eROSITA mission has already produced promising results. We selected 33 INS candidates that can be divided into two soft and hard sub-samples. The 13 soft sources are a generally good match with thermally emitting INSs, though they still require deeper limits in the optical for a secure identification. For the 20 candidates of the hard sub-sample, alternative identifications, such as an AGN or CV nature, may also be viable. We expect that a full characterisation of the selected candidates may increase the number of known XDINSs by one or up to three, a significant increase with respect to the known population of only seven sources.

The next step is to follow up on the candidates and establish their INS identification and characterise their evolutionary state. At X-ray energies, these efforts primarily concern periodicity searches, which are essential to relate newly discovered sources with the several families of Galactic INSs. Likewise, it is important to characterise the neutron star spectrum and atmosphere. For example, the evidence for spectral features, as observed for several thermally emitting INSs \citep[e.g.][]{2013Natur.500..312T,2022A&A...661A..41S,2023IAUS..363..288M,2024A&A...683A.164K}, allows for an estimate of the neutron star magnetic field close to the surface. While such absorption features cannot be identified in shallow \eros\ spectra alone, we note that indirect evidence, such as a large best-fit $\nh$, may be indicative of their presence \citep[see e.g. the case of J13171 in][]{2024A&A...683A.164K}.

Fully characterizing the candidate sample will not only better constrain the population properties of XDINSs but also help in understanding the most common contaminants surviving our search procedure. This promises improvements in future searches since the remaining contaminants may allow for more accurate tailoring of the selection steps towards INSs. In this regard, it is also of high interest to compare our selection with other cross-identification projects, as they are being undertaken within the German \eros\ consortium to select stars, CVs, or generally the most likely cross-match for any \eros\ source \cite[][]{2024A&A...684A.121F,2024A&A...682A..34M}.

\begin{acknowledgements}
We thank the anonymous referee and the editor for helpful suggestions and feedback that improved this paper.

This work was funded by the Deutsche Forschungsgemeinschaft (DFG, German Research Foundation) – 414059771. This work was supported by the project XMM2ATHENA, which has received funding from the European Union's Horizon 2020 research and innovation programme under grant agreement n$^{\rm o}101004168$.

AMP acknowledges the Innovation and Development Fund of Science and Technology of the Institute of Geochemistry, Chinese Academy of Sciences, the National Key Research and Development Program of China (Grant No. 2022YFF0503100), the Strategic Priority Research Program of the Chinese Academy of Sciences (Grant No. XDB 41000000), and the Key Research Program of the Chinese Academy of Sciences (Grant NO. KGFZD-145-23-15).

This work is based on data from eROSITA, the soft X-ray instrument aboard SRG, a joint Russian-German science mission supported by the Russian Space Agency (Roskosmos), in the interests of the Russian Academy of Sciences represented by its Space Research Institute (IKI), and the Deutsches Zentrum für Luft- und Raumfahrt (DLR). The SRG spacecraft was built by Lavochkin Association (NPOL) and its subcontractors, and is operated by NPOL with support from the Max Planck Institute for Extraterrestrial Physics (MPE).

The development and construction of the eROSITA X-ray instrument was led by MPE, with contributions from the Dr. Karl Remeis Observatory Bamberg \& ECAP (FAU Erlangen-Nuernberg), the University of Hamburg Observatory, the Leibniz Institute for Astrophysics Potsdam (AIP), and the Institute for Astronomy and Astrophysics of the University of Tübingen, with the support of DLR and the Max Planck Society. The Argelander Institute for Astronomy of the University of Bonn and the Ludwig Maximilians Universität Munich also participated in the science preparation for eROSITA.

The eROSITA data shown here were processed using the eSASS/NRTA software system developed by the German eROSITA consortium.

For analysing X-ray spectra, we use the analysis software BXA \citep{2014A&A...564A.125B}, which connects the nested sampling algorithm UltraNest \citep{2021JOSS....6.3001B} with the fitting environment XSPEC \citep{1996ASPC..101...17A}.

This research has made use of the VizieR catalogue access tool, CDS, Strasbourg, France (DOI : 10.26093/cds/vizier). The original description of the VizieR service was published in 2000, A\&AS 143, 23.

This research has made use of data obtained from XMMSL2, the Second XMM-Newton Slew Survey Catalogue, produced by members of the XMM SOC, the EPIC consortium, and using work carried out in the context of the EXTraS project ("Exploring the X-ray Transient and variable Sky", funded from the EU's Seventh Framework Programme under grant agreement no. 607452).

This work made use of Astropy:\footnote{http://www.astropy.org} a community-developed core Python package and an ecosystem of tools and resources for astronomy \citep{astropy:2013, astropy:2018, astropy:2022}.

\end{acknowledgements}
\bibliographystyle{aa}
\bibliography{ref_ero_insc_sample}

\begin{appendix}
\end{appendix}
\end{document}